\PassOptionsToPackage{dvipsnames}{xcolor}
\documentclass[preprint2,twocolumn]{aastex701}

\usepackage{mathtools}
\usepackage{todonotes}
\usepackage{xcolor}

\usepackage[T1]{fontenc}
\usepackage[dvipsnames]{xcolor}
\usepackage{graphicx,orcidlink}
\usepackage{amsfonts,amsmath}
\usepackage{graphicx}
\usepackage{newtxtext,newtxmath}
\usepackage[normalem]{ulem}
\usepackage{cancel}
\usepackage{listings}
\usepackage{booktabs}
\usepackage{tabularx}
\usepackage{threeparttable}

\usepackage{hyperref}
\definecolor{purple}{RGB}{128, 0, 128}
\hypersetup{
    colorlinks=true,
    linkcolor=purple,        
    filecolor=green,     
    urlcolor=purple,         
    citecolor=purple        
}

\allowdisplaybreaks
\numberwithin{equation}{section}

\newcommand{\triright}{%
    \mbox{%
        \begin{tikzpicture}[baseline=-0.5ex, line cap=round]
            \draw[line width=0.12ex] (0,0) -- (-0.1, 0.1) 
                                     (0,0) -- (-0.1, -0.1) 
                                     (0,0) -- (0.1, 0.0);
        \end{tikzpicture}%
    }%
}

\makeatletter
\renewcommand*{\@textcolor}[3]{%
  \protect\leavevmode
  \begingroup
    \color#1{#2}#3%
  \endgroup
}
\makeatother

\newcommand{\ab}[1]{\left|#1\right|}

\newcommand{\pa}[1]{\left(#1\right)}

\begin{document}

\title{Bayesian Evidence for Inspiraling Hotspot Motion in the Galactic Center}

\author{Pablo Ruales\,\orcidlink{0000-0002-3199-1025}}
\email[show]{rualpm25@wfu.edu}
\affiliation{Department of Physics, Wake Forest University, Winston-Salem, North Carolina 27109, USA}

\author{Maciek Wielgus\,\orcidlink{0000-0002-8635-4242}} 
\email[]{mwielgus@iaa.es}
\affiliation{Instituto de Astrofísica de Andalucía–CSIC, Glorieta de la Astronomía s/n, E-18008 Granada, Spain}

\author{Delilah E.~A. Gates\,\orcidlink{0000-0002-4882-2674}}
\email[]{delilah.gates@cfa.harvard.edu}
\affiliation{Center for Astrophysics $\arrowvert$ Harvard \& Smithsonian, 60 Garden Street, Cambridge, MA 02138, USA}
\affiliation{Black Hole Initiative at Harvard University, 20 Garden Street, Cambridge, MA 02138, USA}

\author{Alejandro C\'ardenas-Avenda\~no\,\orcidlink{0000-0001-9528-1826}} 
\email[]{cardenas@wfu.edu}
\affiliation{Department of Physics, Wake Forest University, Winston-Salem, North Carolina 27109, USA}

\begin{abstract}
During flaring episodes, polarimetric observations of the Galactic Center black hole Sagittarius A* (Sgr~A*) reveal evolving, loop-like trajectories in the Stokes $Q$--$U$ plane, the so-called $Q$--$U$ loops. Keplerian models of transiently energized orbiting features, hotspots, have been employed to reproduce these polarimetric signatures. In this work, we develop a Bayesian framework to infer the time-dependent kinematics of the hotspot from polarized light curves, using a general prescription for its motion. We model the emitting region's trajectory as a continuous, piecewise sequence of $K$ kinematic segments, each described by radial and azimuthal velocity parameters. We perform analyses with $K=1,2,3$ and compare these models to millimeter-wavelength ALMA observations of Sgr~A* from 2017 April 11 that include two subsequent $Q$--$U$ loops, extending the observational constraints on the kinematics. We examine the effects of the local magnetic field geometry, spectral index, and kinematic profile of the hotspot's trajectory on the polarized signal. We find statistically significant evidence that the emitting region follows a non-geodesic inspiraling trajectory, and this preference over a strictly Keplerian orbit strengthens sharply when a longer data segment is modeled. The inspiraling non-Keplerian models are preferred to a purely Keplerian model based on both the Bayesian log-evidence comparison and the reduced chi-square statistic. Our results offer new insight into the nature of flaring events in the Galactic Center and the episodically produced hotspots associated with them.
\end{abstract}

\section{Introduction}

The supermassive black hole at the center of our galaxy, Sagittarius~A* (Sgr~A*), with mass $M = 4.3\times10^6\,M_\odot$~\citep{GRAVITY:2021xju}, is the closest supermassive black hole to Earth and therefore an ideal target for detailed studies of relativistic accretion. Over the past decades, measurements at X-ray, infrared, and millimeter wavelengths have shown ``flaring'' states, in which the observed radiative flux exhibits enhanced magnitude and variability in comparison to quiescent periods~\citep{GRAVITY:2020lpa,Zamaninasab:2009df,Haggard:2019mro,Mossoux:2020ddc,EventHorizonTelescope:2022ago,Baganoff:2001kw,Genzel:2003as}. Transiently energized hotspots, localized regions of enhanced emission, have been proposed to explain these observations~\citep{1992A&A...257..531K,Broderick:2005my,Broderick:2005jj,Abuter:2018uum,Ball:2020jup,Wielgus:2026qft}. 
On 2017 April 11, the Atacama Large Millimeter/submillimeter Array (ALMA) observed Sgr~A* during the Event Horizon Telescope campaign~\citep{EventHorizonTelescope:2022ago}. The observation followed an X-ray flare~\citep{EventHorizonTelescope:2022apq,Haggard:2019mro} and indicated a clockwise looping structure in the Stokes $Q$--$U$ plane of linear polarization~\citep{Wielgus:2022heh}, the so-called ``$Q$--$U$ loops.'' The data exhibit a large primary loop and a subsequent smaller inner one. 
In recent years, semi-analytical models assuming a hotspot in a fixed circular orbit have been used to explain such observational signatures~\citep{Wielgus:2022heh, GRAVITY:2023avo,Yfantis:2023wsp,Yfantis:2024eab,Levis:2023tpb,Tlemissov:2026kuu}. For that particular observation, the models infer an emitter at $r\sim 9$--$13\,M$\footnote{We use geometric units $G = c = 1$ and assume a timescale $GM/c^3 = 21.2$\,s~\citep{GRAVITY:2021xju} for Sgr~A*.} and inclination $\theta_{\rm obs} \sim 160^\circ$~\citep{Wielgus:2022heh,Yfantis:2023wsp,Levis:2023tpb}. Such constraints, alongside studies of the implied kinematics of the bright feature, are very important for characterization of the Sgr~A* system, and more broadly for studies of magnetized accretion, its dynamics, and the nature of plasma dissipation~\citep{Porth:2020txf, Najafi-Ziyazi:2023oil, Ricarte:2025aix}.

\begin{figure*}[]
    \centering
        \includegraphics[width=\textwidth]{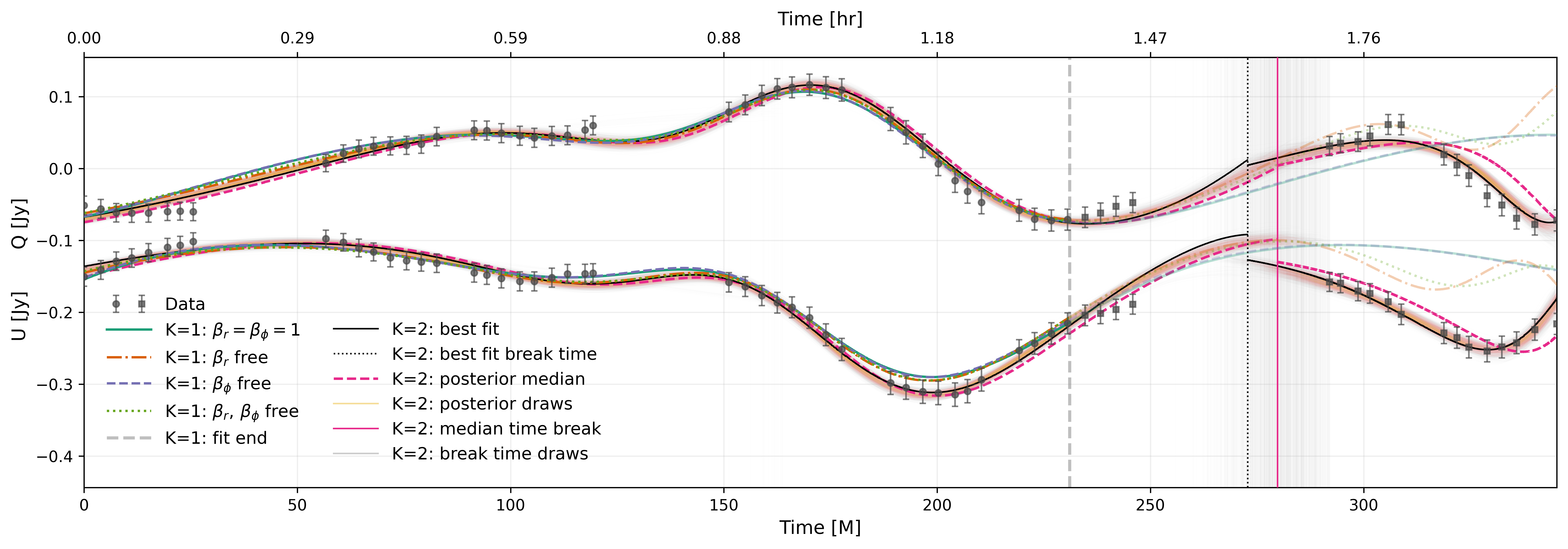}%
        
    \caption{Observed and simulated Stokes parameters $Q(t)$ (upper lines) and $U(t)$ (lower lines) for the $K=1$ models and the $E_2$ model of Table~\ref{tab:fit_results}. The data points correspond to the 2017 April 11 ALMA observation~\citep{Wielgus:2022heh}, with uncertainties computed using Eq.~\ref{eq:error}, drawn as circles within the range fitted by the $K=1$ models, $t_{\rm obs}\sim230\,M$ (gray dashed vertical line), and as squares beyond it. The lower horizontal axis indicates time in geometric units $[M]$, which has been converted from hours, shown on the upper axis, assuming a black hole mass of $M=4.3\times10^6\,M_\odot$. For the $K=1$ family, model realizations are plotted with solid colors up to that line and with increased transparency thereafter, where the models are continued beyond the fitted range and their agreement with the remaining data degrades. Posterior samples of the inferred break time for the $E_2$ model are indicated by vertical light-gray lines, with the associated $Q$ and $U$ trajectories shown in light gold. The best-fitting realization of the highest-evidence model, $E_2$ ($K=2$, $\beta_r$ and $\beta_\phi$ free, single $\alpha_\nu$), is shown in black, while the posterior median is shown in fuchsia, with the best-fit and median break times marked by the black dotted and fuchsia solid vertical lines, respectively.}
    \label{fig:K1vsK2}
\end{figure*}

Although the inferred orbital radii are close to the black hole, the spin is a subdominant parameter at those radii, and the details of the magnetic field configuration and the kinematic profile are more relevant for the observables~\citep{Gelles:2021kti,Wielgus:2022heh,Ruales:2026xjb}. In particular, the aforementioned models suggest a local magnetic field with a dominant vertical component (perpendicular to the accretion disk). While the overall structure of these polarimetric signatures is a consequence of the background axisymmetry and even some finer features are of a special-relativistic origin~\citep{Vincent:2023sbw}, a curved-spacetime treatment is required to reproduce the details of the observable morphology~\citep{Gelles:2021kti,Vos:2022yij,Vincent:2023sbw,Ruales:2026xjb}. This is because the observed signal is distorted by strong-field effects such as gravitational lensing and gravitational redshift, which modify the shape and asymmetry of $Q$--$U$ loops as the hotspot's trajectory develops. To date, only the primary outer $Q$--$U$ loop reported in \citet{Wielgus:2022heh}, spanning around 80 minutes of observations ($t_{\rm obs}\sim 230\,M$, up to the gray dashed vertical line in Fig.~\ref{fig:K1vsK2}), has been fitted successfully with a hotspot model~\citep{Wielgus:2022heh,Yfantis:2023wsp}, since the later portion of the observation, which contains the second loop, is inconsistent with any periodic trajectory. Prior fits to this dataset have also tested for deviations from Keplerian motion within the circular-orbit framework~\citep{Wielgus:2022heh,Yfantis:2023wsp,Levis:2023tpb}.

In this work we fit up to $\approx 2$ hrs ($t_{\rm obs}\sim 350\,M$) of the 2017 observation (see Fig.~\ref{fig:K1vsK2}), which includes a portion of the inner loop that appears after the primary one, using a more general, non-circular model of motion. We use the non-Keplerian inspiral model introduced in \citet{Ruales:2026xjb}. We model the emitting region's trajectory as a continuous, piecewise sequence of $K$ kinematic segments, each described by radial and azimuthal velocity parameters defined relative to the circular-geodesic (Keplerian) values. The inference is performed using a parallel-tempered Markov chain Monte Carlo (PT-MCMC) algorithm, and we compare different kinematic profiles via Bayesian evidence. Because these velocity profiles change discontinuously between segments, while the trajectory is always continuous, the observed polarimetric curve has jumps at the break times, which are marked in Fig.~\ref{fig:K1vsK2} by the vertical light-gray lines.

The single-segment $(K=1)$ fit matches the kinematic prescriptions considered in prior work~\citep{Wielgus:2022heh,Yfantis:2023wsp}, in that the angular velocity parameter is broadly consistent with Keplerian motion but not tightly constrained. When we allow this segment to have a radial component, that is, an inspiraling motion, the data prefer a mild inspiral for the primary loop. However, when the inner loop is included, the fit quality rapidly deteriorates (see the light-colored $K=1$ curves beyond the gray-dashed vertical line in Fig.~\ref{fig:K1vsK2}). This result motivates allowing the kinematics to change along the trajectory. 

When considering two segments $(K=2)$, we find that the posteriors of the second segment are consistent with a faster inspiral beyond where previous work has finished the fit (see the gray-dashed vertical line, which denotes the end of the fitting in \citet{Yfantis:2023wsp}, and light-gray lines, the break times inferred by our analysis). Moreover, an inferred break time later than the maximum time considered in \citet{Yfantis:2023wsp} suggests that a single-segment analysis is blind to the non-Keplerian evolution of the hotspot kinematics. Therefore, the small preference for an inspiral found within $t_{\rm obs}\sim 230\,M$ is expected, since a single segment can only recover a kinematic profile averaged over a window that excludes the break. Our results suggest that the emitter follows an inspiraling trajectory with a growing magnitude of its radial velocity as the trajectory develops, see the trajectory corresponding to the median of the posterior of model $E_2$ in Fig.~\ref{fig:K2_QU_traj_nonkep}. We also perform a $K=3$ analysis and find that a third segment for the data shown in Fig.~\ref{fig:K1vsK2} is disfavored over the two-segment fit. 

The structure of the rest of this paper is as follows. In Sec.~\ref{sec:theory}, we provide the theoretical background required to compute the polarized emission from inspiraling emitters in the Kerr geometry. In Sec.~\ref{sec:data_analysis} we present the setup for the PT-MCMC algorithm used to fit the observational data, together with the results of the model and a Bayesian evidence comparison across model variants. Lastly, in Sec.~\ref{sec:conclusions} we discuss our results and their physical interpretation.

\section{Inspiraling Hotspot Model}\label{sec:theory}

Using the inspiraling hotspot model from \citet{Ruales:2026xjb}, we build a pipeline that computes the polarized emission of an inspiraling emitter on a generalized, not necessarily circular or geodesic, trajectory in the Kerr spacetime. In this section, we briefly review two structural ingredients of the model: the kinematics of an equatorial emitter and the polarization of the emission as observed on a screen at infinity. For more details we refer the reader to \citet{Ruales:2026xjb}.

\subsection{Hotspot Kinematics}
\label{sec:hotspot_kinematics}

The model proposed in \citet{Ruales:2026xjb} uses a generalized flow in the equatorial plane around a Kerr black hole described by the mass $M$ and dimensionless spin parameter $a$~\citep{Pu:2016qak,Vincent:2022fwj,Cardenas-Avendano:2022csp}. The velocity field is an interpolation between a generalized Keplerian flow first defined by \citet{Cunningham:1975zz} and a zero angular momentum geodesic free-fall from infinity, with the interpolation controlled by two dimensionless parameters $\beta_r$ and $\beta_\phi$. We refer to these parameters collectively as the ``Keplerianity parameters,'' where $\beta_r$ ($\beta_\phi$) sets the interpolation between the Keplerian radial (angular) velocity profile and that of a particle released from rest at infinity with vanishing angular momentum. Setting $\beta_r=\beta_\phi=1$ recovers Cunningham's disk model, which describes a stable circular orbit outside the innermost stable circular orbit (ISCO) and a geodesic plunge inside it. Values of $\beta_r<1$ introduce a radial component of the flow, allowing for inspiraling trajectories, and outspiraling for $\beta_r>1$, outside the ISCO, and $\beta_r=0$ corresponds to the geodesic radial infall velocity profile. Values of $\beta_\phi<1$ reduce the angular velocity below the Keplerian (circular-geodesic) value at the same radius, describing a flow whose rotational support is partially removed, and $\beta_\phi=0$ corresponds to a flow with vanishing angular momentum. The parametrized four-velocity is
\begin{equation}
    \tilde u=\tilde u^t\left(\partial_t-\tilde\iota\,\partial_r+\tilde\Omega\,\partial_\phi\right),
\end{equation}
where $\tilde{\iota} = - \tilde{u}^r/\tilde{u}^t$ and $\tilde{\Omega} = \tilde{u}^\phi/\tilde{u}^t$ are the radial and angular velocities, respectively. The negative sign in $\tilde{\iota}$ makes it positive for infalling motion. The interpolation is performed on the radial four-velocity component and on the angular velocity
\begin{equation}   
    \begin{split}        \tilde{u}^r&=\hat{u}^r+\pa{1-\beta_r}\pa{\bar{u}^r-\hat{u}^r}, \\ \tilde{\Omega}&=\hat{\Omega}+\pa{1-\beta_\phi}\pa{\bar{\Omega}-\hat{\Omega}},
    \end{split}
\end{equation}
where hatted quantities denote the Keplerian values and barred quantities those of a particle released from rest at infinity, and $\tilde u^t$ then follows from the normalization $\tilde u\cdot \tilde u=-1$. This velocity field is implemented in the adaptive analytical ray-tracing code \texttt{AART}~\citep{Cardenas-Avendano:2022csp}. For a fixed spacetime geometry, fully described by the spin $a$ and the observer's inclination $\theta_{\rm obs}$, \texttt{AART} computes the map from every screen pixel with Bardeen coordinates $(\alpha,\beta)$ (defined in Sec.~\ref{sec:observed_polarization}) in lensing band $n$ to its source-plane coordinates $(r,\phi)$ and the photon travel time. 

The finite light travel time between an emitter moving in a time-dependent trajectory and the observer is accounted for with a physically correct slow-light treatment~\citep{Rojas-Paternina:2026nsq}. The observer can simultaneously receive light emitted at different times from different points along the orbit, thus modifying the morphology of the observed emission. The slow-light correction maps the hotspot emission time to the photon arrival time at the observer's screen, for every active screen pixel of the direct image $(n=0)$. Higher-order images ($n\ge1$) are strongly demagnified relative to the direct image and are not included in the fits presented in this work. The backward ray-tracing routine records a Boyer--Lindquist coordinate time $t_{\rm ray}^{(n)}$. This is the light travel time corresponding to a single photon arriving on the screen, defined up to an additive constant set by the arbitrary observer distance from the source. Only the differences of $t_{\rm ray}^{(n)}$ between pixels are physically meaningful, so we define the zero point of relative delay as the trajectory's own first sample point, $t=0$, giving the observer arrival time for each point along the trajectory
\begin{equation}\label{eq:tobs0}
    t_{\rm obs}^{(0)}(t)=t_{\rm emit}(t)+\left[t_{\rm ray}^{(0)}(0)-t_{\rm ray}^{(0)}(t)\right].
\end{equation}
By construction $t_{\rm obs}^{(0)}(0)=t_{\rm emit}(0)$, so the first trajectory point defines zero relative delay, and every later point is advanced or delayed in observer time according to how much shorter or longer its null geodesic is relative to the first point's. Since we use a backward ray-tracing algorithm, the recorded light travel times are negative, so a trajectory point whose ray takes longer to reach the screen has a more negative $t_{\rm ray}^{(0)}(t)$, the bracket in Eq.~\ref{eq:tobs0} is positive, and $t_{\rm obs}^{(0)}(t)$ is correctly shifted to a later observer time.

\subsection{Observed Polarization}
\label{sec:observed_polarization}

The polarized emission is calculated in a local frame co-moving with the fluid and then transformed to the global frame of the black hole. This connection, as detailed in \citet{Ruales:2026xjb}, requires an orthonormal tetrad $e'^\mu{}_{(a)}$ built in two steps. We first construct the zero-angular-momentum observer (ZAMO) orthonormal tetrad $e^\mu{}_{(a)}$~\citep{Bardeen:1972fi}, which is fixed by the Kerr metric in Boyer--Lindquist coordinates and sets up a locally flat frame at the emission point. This tetrad is then boosted into the fluid frame along the emitter's three-velocity $\vec{\mathcal V}$, the projection of the four-velocity $\tilde u^\mu$ onto the ZAMO frame. The synchrotron polarization vector in the fluid frame, $\vec{f'}=\vec{p'}\times\vec{B'}/|\vec{p'}|$~\citep{Rybicki:2004hfl}, is perpendicular to both the local photon momentum $\vec {p'}$ and the local magnetic field $\vec{B'}=(B'_r,B'_\theta,B'_\phi)$, and is transformed back to the Boyer--Lindquist frame with the same fluid-frame tetrad. Once the polarization four-vector is known in the global frame of the black hole, its parallel transport to the observer is encoded in the Penrose--Walker constant~\citep{Walker:1970un}, which in its equatorial form $(\theta=\pi/2)$ is
\begin{equation}
    \kappa=\kappa_1+i\kappa_2=r\,(\mathcal{P}_{\mathcal A}-i\mathcal{P}_{\mathcal B}).
\end{equation}
Because $\kappa$ is conserved along null geodesics in Kerr spacetime, it can be evaluated once at emission and used directly to reconstruct the polarization as seen by a distant observer, with $\mathcal{P}_{\mathcal A}$ and $\mathcal{P}_{\mathcal B}$ built from the photon momentum and polarization four-vector components.

A photon reaching the observer's screen at Bardeen coordinates $(\alpha,\beta)$ satisfies $\alpha=-\lambda/\sin\theta_{\rm obs}$, $\beta=\pm\sqrt{\eta+a^2\cos^2\theta_{\rm obs}-\lambda^2\cot^2\theta_{\rm obs}}$~\citep{Bardeen:1972fi}, where $\lambda$ and $\eta$ are the photon's energy-rescaled angular momentum and Carter constant. Its electric vector position angle (EVPA), $\chi$, follows directly from $\kappa$ and $(\alpha,\beta)$~\citep{Gelles:2021kti},
\begin{equation}
    \chi=\arctan\left(\frac{\omega\kappa_1-\beta\kappa_2}{\beta\kappa_1+\omega\kappa_2}\right), \quad \omega\equiv-(\alpha+a\sin\theta_{\rm obs}).
\end{equation}
The screen-frame polarization components are
\begin{equation}
    (f_{[\alpha]},f_{[\beta]})=(\beta\kappa_2-\omega\kappa_1,\,\beta\kappa_1+\omega\kappa_2)/(\beta^2+\omega^2),
\end{equation}
and the observed polarization $(f^{\rm obs}_{[\alpha]},f^{\rm obs}_{[\beta]})$ is then obtained by scaling the transported polarization $(f_{[\alpha]},f_{[\beta]})$ by a power of the redshift factor $g\equiv\nu_{\rm o}/\nu_{\rm e}$ set by the spectral index $\alpha_\nu$. We adopt the flux convention $S_\nu\propto\nu^{-\alpha_\nu}$, so that optically thin synchrotron emission corresponds to $\alpha_\nu>0$. The scaling of the observed intensity with $g$ then follows from the invariance of $I_\nu/\nu^3$ along a null geodesic, i.e., $I_{\nu_{\rm o}}(\nu_{\rm o})=g^3\,I_{\nu_{\rm e}}(\nu_{\rm e})$,
combined with the optically thin emission spectrum evaluated at the emitted frequency $\nu_{\rm e}=\nu_{\rm o}/g$, $I_{\nu_{\rm e}}(\nu_{\rm e})\propto\nu_{\rm e}^{-\alpha_\nu}\propto g^{\alpha_\nu}\nu_{\rm o}^{-\alpha_\nu}$, and therefore
\begin{equation}\label{eq:fluxscaling}
    I_{\nu_{\rm o}}(\nu_{\rm o})\ \propto\ g^{3+\alpha_\nu}.
\end{equation}
Since the Stokes parameters are quadratic in $\vec f$ (i.e., $|\vec f\,|^2\propto I_\nu$), the corresponding scaling of $\vec f$ is the square root of Eq.~\ref{eq:fluxscaling}, as used in \citet{Ruales:2026xjb} and \citet{Gelles:2021kti},
\begin{equation}\label{eq:redshift}
    \big(f^{\rm obs}_{[\alpha]},f^{\rm obs}_{[\beta]}\big)=g^{(3+\alpha_\nu)/2}\big(f_{[\alpha]},f_{[\beta]}\big),
\end{equation}
so that the Stokes parameters are~\citep{EventHorizonTelescope:2021btj}
\begin{equation}
    Q=\big(f^{\rm obs}_{[\beta]}\big)^2-\big(f^{\rm obs}_{[\alpha]}\big)^2, \quad
    U=-2\,f^{\rm obs}_{[\alpha]}f^{\rm obs}_{[\beta]},
    \label{eq:QUdef}
\end{equation}
which, by Eq.~\ref{eq:redshift}, scale as $Q,U\propto g^{3+\alpha_\nu}$.

The magnetic field $\vec{B'}$ and the spectral index $\alpha_\nu$, together with the kinematic parameters of Sec.~\ref{sec:hotspot_kinematics}, are the free parameters of the fit, as shown in Eq.~\ref{eq:params}. We implement this framework, which computes the polarized emission along a non-geodesic inspiraling trajectory, within a PT-MCMC algorithm using \texttt{reddemcee}~\citep{2026A&A...706A.323P} to sample the posterior distribution of the model parameters and to estimate the Bayesian evidence for the fitting of the 2017 April 11 ALMA dataset.

\section{Bayesian Analysis}\label{sec:data_analysis}

In this section we present the configuration of a PT-MCMC algorithm in which the model from Sec.~\ref{sec:theory} is implemented, followed by the results summarized in Fig.~\ref{fig:K1vsK2}, and shown in Table~\ref{tab:fit_results} in detail.

\subsection{Parallel-Tempered MCMC}
\label{sec:ptMCMC}

The observed polarized emission from an inspiraling hotspot calculated with the model described in Sec.~\ref{sec:theory} is compared to ALMA data using a PT-MCMC algorithm. The hotspot's trajectory is segmented into a sequence of up to $K$ continuous kinematic segments. The number of segments is fixed prior to the start of the analysis. A run with $K=1$ would contain a single set of kinematic parameters throughout the trajectory, while runs with $K>1$ allow the radial and azimuthal components of the velocity to evolve between time intervals. This algorithm provides a controlled way of testing how much kinematic change is required by the data. Additionally, we can fix one or more of the kinematic parameters (e.g., $\beta_r=1$) for the duration of a run and the dimensionality of the problem will reduce accordingly. We can also sample a single $\alpha_\nu$ throughout the kinematic segments for one of the experiments, instead of sampling one per segment. Therefore, a complete analysis would require parallel runs with several values of $K$ and different model configurations, to later assess which model the data prefer.

For a chosen value of $K$, the sampled parameter vector is
    \begin{equation}\label{eq:params}
        \begin{split}
        \theta_K=\{&r_{\rm init},\phi_{\rm init},
    \beta_{r,1},\ldots,\beta_{r,K},
    \beta_{\phi,1},\ldots,\beta_{\phi,K},\\
    &\tau_1,\ldots,\tau_{K-1},
    Q_{{\rm sha},1},\ldots,Q_{{\rm sha},K},\\
    &U_{{\rm sha},1},\ldots,U_{{\rm sha},K},\text{PA},\\
    &B_r',B_\theta',B_\phi',\alpha_1,\ldots,\alpha_K\},
        \end{split}
    \end{equation}
and the dimensionality of the fit is $N_{\rm dim}=6K+5$, or $N_{\rm dim}=5K+6$ when a single $\alpha_\nu$ is shared across segments. The initial position of the trajectory is given by $(r_{\rm init},\phi_{\rm init})$, and the Keplerianity parameters $(\beta_{r,i},\beta_{\phi,i})$ of Sec.~\ref{sec:hotspot_kinematics} set the kinematics of each segment. The time variables $\tau_i$ determine the relative weights of the $K$ time intervals through a unity-normalized mapping that sets the segment durations. A constant offset in the $Q$--$U$ plane per segment, $(Q_{\rm sha},U_{\rm sha})$, represents the background emission, as in \citet{Wielgus:2022heh} and \citet{Yfantis:2023wsp}. A single global rotation angle, PA, representing the overall orientation of the observer's sky frame relative to the coordinate frame, is applied to the model's Stokes plane (Eq.~\ref{eq:evparot}), and $\vec{B'}=(B_r',B_\theta',B_\phi')$ is the local magnetic field of Sec.~\ref{sec:observed_polarization}. Lastly, $\alpha_i=(3+\alpha_{\nu,i})/2$ is the redshift exponent of Eq.~\ref{eq:redshift}, from which the spectral index follows. Returning to the time variables, only $K-1$ of them are free, with the last one defined to be zero, preventing degeneracies in the time intervals,
\begin{equation}
    \vec{\tau}=\{\tau_1,\tau_2,\ldots,\tau_{K-1},0\}.
\end{equation}
The normalized time weights are defined as
\begin{equation}\label{eq:time_weights}
    w_i \equiv \frac{\exp({\tau}_i)}{\sum_{m=1}^{K}\exp({\tau}_m)},
    \qquad
    \sum_{i=1}^{K}w_i=1 ,
\end{equation}
and the physical duration of the segment $i$ is calculated from its weight $w_i$ as
\begin{equation}
    \Delta t_i = \Delta t_{\rm min}+ \left(T_{\rm sim}-K\Delta t_{\rm min}\right)w_i,
    \label{eq:fixedK_duration}
\end{equation}
where $T_{\rm sim}=t_{\rm max} - t_{\rm min}$ is the total duration of the simulated trajectory, set equal to the span of the observed data after shifting the first observed time to zero, and $\Delta t_{\rm min}$ is a small minimum segment duration. This guarantees that every active segment has positive duration and that the segments exactly cover the simulated interval
\begin{equation}
    \sum_{i=1}^{K}\Delta t_i = T_{\rm sim}.
\end{equation}
Therefore, $\tau_i$ represents a relative duration, where large positive values make the corresponding segment longer, large negative values make it shorter, and equal values produce approximately equal segment durations.

\begin{table}[]
    \centering
    \caption{Prior distributions for all sampled parameters, for a fixed-$K$ model with $K$ kinematic segments. $r_{\rm ms}(a)\approx 2.51\,M$ is the Kerr ISCO radius at the fixed spin $a=0.87$. All priors are uniform on the stated support and are normalized, so that the prior volume is accounted for in the evidence comparison.}
    \label{tab:priors}
    \begin{threeparttable}
    \setlength{\tabcolsep}{4pt}
    \renewcommand{\arraystretch}{1.3}
    \begin{tabular}{llc}
        \toprule
        \multicolumn{2}{c}{Parameter} & Support\\
        \midrule
        Initial radius        & $r_{\rm init}$   & $[1.5\,r_{\rm ms}(a),\,17.0]\,M$ \\
        Initial azimuth       & $\phi_{\rm init}$& $[0,\,2\pi)$ \\
        Radial Keplerianity   & $\beta_{r,i}$    & $[0,\,1.2]^K$\tnote{a} \\
        Angular Keplerianity  & $\beta_{\phi,i}$ & $[0,\,1.2]^K$\tnote{a} \\
        Duration weights      & $\tau_i$, $i<K$  & flat in duration\tnote{b} \\
        Background $Q$   & $Q_{{\rm sha},i}$& $[-1,\,1]^K$ Jy \\
        Background $U$   & $U_{{\rm sha},i}$& $[-1,\,1]^K$ Jy \\
        Sky rotation          & $\text{PA}$      & $[-90,\,90)^\circ$ \\
        Radial field          & $B_r'$            & $[0,\,1]$ \\
        Poloidal field        & $B_\theta'$       & $[0,\,1]$ \\
        Toroidal field        & $B_\phi'$         & $[0,\,1]$ \\
        Redshift exponent     & $\alpha_i$       & $[0.5,\,5.0]^{K}$ \\
        \bottomrule
    \end{tabular}
    \begin{tablenotes}\footnotesize
    \item[a] The $K$ Keplerianity values are additionally restricted to be monotonic, in either direction. This has no effect for $K\leq2$ and reduces the prior volume by $K!/2$ for $K\geq3$.
    \item[b] The $K-1$ sampled $\tau_i$ are mapped through the softmax of Eq.~\ref{eq:time_weights} to duration fractions $w_i$ on the $(K-1)$-simplex. The prior is flat in physical duration subject to $\sum_i\Delta t_i=T_{\rm sim}$ and $\Delta t_i\geq\Delta t_{\rm min}$. No duration is sampled for $K=1$.
    \end{tablenotes}
    \end{threeparttable}
\end{table}

The polarimetric signal is anti-symmetric with respect to the equatorial plane, in the sense that the Stokes $Q$ and $U$ observed at $\theta_{\rm obs}<90^\circ$ are equivalent to $Q$ and $-U$ observed at $\theta_{\rm obs}>90^\circ$ from the same signal. We therefore simulate at $\theta_{\rm obs}=20^\circ$ and multiply the computed Stokes $U$ by $-1$, which reproduces the signal as observed from below the equatorial plane at $\theta_{\rm obs}\sim160^\circ$, matching the inclination inferred in previous work~\citep{Wielgus:2022heh}.

For each proposed parameter vector, the hotspot's motion is propagated through the $K$ segments sequentially. The final position of each segment $i$ is taken by $i+1$ as its initial conditions, making the trajectory physically connected. For every segment, the Stokes parameters $Q$--$U$ are computed along the trajectory, and are immediately rescaled by the lensing magnification of the hotspot at each trajectory point $\mu_{\rm proper}(r)$, derived in Appendix~\ref{sec:magnification} following \citet{Zhu:2026jom}. Then they are concatenated into a single piecewise curve, the sign of $U$ is flipped as described above, and the entire curve is rescaled by a single global amplitude factor $s(\theta_K)$ that matches the size of the model's $Q$--$U$ loop to that of the data. To calculate $s(\theta_K)$, a Taubin fit~\citep{Taubin1991} is performed separately on the data and on the model's $Q$--$U$ track, and $s(\theta_K)$ is the ratio of the two fitted radii. The loop radius is set jointly by the kinematics and by the intrinsic emissivity of the hotspot, which our model does not predict in absolute units, so the two are inseparable and the radius alone cannot constrain the trajectory. Rescaling absorbs this degeneracy into a single deterministic factor rather than a sampled parameter, leaving the trajectory parameters to be determined by the shape and timing of the track. After this amplitude calibration, each point is shifted by the per-segment background offset $(Q_{{\rm sha},i},U_{{\rm sha},i})$ of the segment it belongs to.

Lastly, a global rotation angle, PA, of the entire $Q$--$U$ plane is also inferred. This angle absorbs both the orientation of the observer's sky frame and any external Faraday rotation, two contributions that this model cannot distinguish~\citep{Wielgus:2023akf,Ricarte:2025aix}. This follows previous studies, e.g., \citet{Wielgus:2022heh} included such an angle as a correction to a model similar to the one presented in this work, and \citet{Yfantis:2023wsp} include this parameter in their own analysis. Given that the EVPA is defined as
\begin{equation}\label{eq:evpa}
\chi=\frac{1}{2} \arg \left( Q+ \mathrm{i} U \right),
\end{equation}
we in turn define the rotation of the $Q$--$U$ plane as
\begin{equation}\label{eq:evparot}
    \begin{pmatrix}Q'\\U'\end{pmatrix}=
    \begin{pmatrix}\cos(2\,\text{PA}) & -\sin(2\,\text{PA})\\ \sin(2\,\text{PA}) & \cos(2\,\text{PA})\end{pmatrix}
    \begin{pmatrix}Q\\U\end{pmatrix}.
\end{equation}
Since $Q_{\rm sha}$ and $U_{\rm sha}$ are fitted before this rotation is applied, the inferred $Q$--$U$ background offset has to be rotated by Eq.~\ref{eq:evparot} to obtain the value as observed at the screen.

For all experiments conducted in this study, we assign uniform (flat) prior distributions to all model parameters, as summarized in Table~\ref{tab:priors}. Once the priors are set for the entirety of the run, the algorithm calculates the Stokes $Q$--$U$ values within the parameter space, starting from an initial guess. The initial vector is heuristically determined such that it would produce a trajectory that fits within the time frame of the dataset. Then, the computed $Q$--$U$ values are compared to the data points through a Gaussian likelihood, closely following \citet{Yfantis:2023wsp}, 
\begin{equation}\label{eq:likelihood}
    \begin{split}
        \ln \mathcal{L}(\theta_K) = -\frac{1}{2}\sum_j \left( \frac{\left[ Q_j - \hat{Q}_j(\theta_K) \right]^2}{\sigma_{Q,j}^2} + \ln\!\big(2\pi\sigma_{Q,j}^2\big) \right.\\
+ \left. \frac{\left[ U_j - \hat{U}_j(\theta_K) \right]^2}{\sigma_{U,j}^2} + \ln\!\big(2\pi\sigma_{U,j}^2\big) \right),
    \end{split}
\end{equation}
\begin{equation}\label{eq:error}
\sigma_{Q,j}=\sigma_{U,j}=\sigma_{j} = \epsilon_s \sqrt{\hat{Q}_j^2+\hat{U}_j^2} + \epsilon_t \ ,
\end{equation}
where $\epsilon_s=0.02$ and $\epsilon_t=0.01$\,Jy are the fractional systematic error and the thermal noise, respectively, adopted from \citet{Yfantis:2023wsp}. Inside the likelihood this is evaluated at the model's own prediction, $\sigma_{Q,j}=\sigma_{U,j}=\sigma\big(\hat Q_j(\theta_K),\hat U_j(\theta_K)\big)$, so the uncertainty assigned to each data point scales with the polarized flux the model itself predicts, and the $\ln(2\pi\sigma^2)$ terms in Eq.~\ref{eq:likelihood} depend on $\theta_K$. The error bars shown in the figures use the same expression evaluated at the observed values, $\sigma(Q_j,U_j)$.

Equation~\ref{eq:likelihood} is the logarithm of a Gaussian likelihood, and it separates into two parts,
\begin{equation}
        \ln \mathcal{L}(\theta_K) = -\frac{1}{2} \chi^2(\theta_K) -\frac{1}{2}\sum_j \left[\ln\big(2\pi\sigma_{Q,j}^2\big)+\ln\big(2\pi\sigma_{U,j}^2\big)\right],
\end{equation}
where the first term is built from the sum of squared differences between data and model, each weighted by its own uncertainty,
\begin{equation}\label{eq:chi2}
    \chi^2(\theta_K) \equiv  \sum_j \left( \frac{\left[ Q_j - \hat{Q}_j(\theta_K) \right]^2}{\sigma_{Q,j}^2} + \frac{\left[ U_j - \hat{U}_j(\theta_K) \right]^2}{\sigma_{U,j}^2} \right),
\end{equation}
and the second is the Gaussian normalization. We report $\chi^2$ on its own as a goodness-of-fit diagnostic. Because $\sigma$ depends on the model prediction, the second term is not the same at every $\theta_K$, so $\chi^2(\theta_K)\neq-2\ln\mathcal{L}(\theta_K)$. We evaluate the model at the highest-posterior sample recovered by the post-burn-in cold chain, following the same convention as \citet{Yfantis:2023wsp}. We refer to this sample as the best fit throughout this work. To compare fits of different dimensionality on a common scale, we define an effective $\chi^2$
\begin{equation}\label{eq:chi2eff}
    \chi^2_{\rm eff} = \frac{\chi^2}{n_d-n_f},
\end{equation}
where $n_d=2N$ is the total number of data points ($N$ points per polarization track, with $Q$ and $U$ counted separately, hence the factor $2$) and $n_f=N_{\rm dim}$ is the number of sampled parameters (Eq.~\ref{eq:params}) for the model variant being evaluated. We also calculate the Bayesian evidence to compare results from different variants of the model, using two methods already integrated within \texttt{reddemcee}~\citep{2026A&A...706A.323P}. First, we calculate the evidence via thermodynamic integration $(\ln Z_{\rm ti})$, which integrates the average log-likelihood (Eq.~\ref{eq:likelihood}), evaluated at each rung of the temperature ladder, over inverse temperature. Second, we use the stepping-stones method $(\ln Z_{\rm ss})$, a discrete sum over the temperature ladder that accumulates the ratio of likelihood-weighted volumes between adjacent rungs, from the prior ($T\to\infty$) to the posterior ($T=1$). In the limit of an infinitely dense temperature ladder both estimators are equivalent. For the full expressions and a more in-depth explanation we refer the reader to \citet{2026A&A...706A.323P}. Since $\ln Z_{\rm ss}$ is built from ratios between adjacent rungs rather than from a quadrature along the whole ladder, it does not accumulate the discretization error of the integral, and its uncertainty reflects only the Monte Carlo variance of each ratio. Across our runs this makes $\ln Z_{\rm ti}$ the more affected of the two by multi-modal posteriors and phase transitions, where the average log-likelihood changes rapidly between rungs. This is reflected in the uncertainties shown in Table~\ref{tab:fit_results}, which are typically smaller for $\ln Z_{\rm ss}$ than for $\ln Z_{\rm ti}$ in every run reported in this work.

As a convergence metric we use the autocorrelation time $\tau$ computed by \texttt{reddemcee} for each sampled parameter. The highest autocorrelation time from each experiment is used to thin the posterior. We test that the Bayesian evidence remains consistent between the full and thinned chains, which supports that the invariant distribution has been reached. Every result shown in Table~\ref{tab:fit_results} passes this test, and an effective sample size (ESS) is computed for each case, $\text{ESS}=(N_{\rm walkers}\cdot N_{\text{post-burn-in-steps}})/\tau$, as a measure of the number of effectively independent samples.

With the Bayesian structure in place, we fit the observed Stokes $Q(t),U(t)$ parameters of the 2017 April 11 ALMA $229$~GHz observation of Sgr~A* reported in \citet{Wielgus:2022heh}, which immediately followed an X-ray flare~\citep{EventHorizonTelescope:2022ago,Haggard:2019mro}. Following \citet{Yfantis:2023wsp}, we sub-sample the original $4$~s cadence data at the observing frequency of 229\,GHz onto a fixed $80$~s time grid starting at the first observed point. For each grid node we retain the single raw sample nearest to it, provided one lies within $40$~s, and nodes with no such sample are dropped. The $80$~s spacing reduces the time correlation between adjacent points relative to the raw cadence while retaining enough samples that the loop morphology is preserved. In Sec.~\ref{sec:K1results} we use a subsample truncated at $t_{\rm obs}\sim230\,M$, and in Sec.~\ref{sec:K2_results} a subsample truncated at $t_{\rm obs}\sim350\,M$.

Given the weak dependence of the $Q$--$U$ loops on the spin parameter at the radii of the primary loop~\citep{Wielgus:2022heh,Gelles:2021kti}, all analyses presented here adopt a fixed geometric configuration with spin parameter $a=0.87$, taken from the posterior of the non-Keplerian model in \citet{Yfantis:2023wsp}. For simplicity, and in accordance with prior studies of this dataset~\citep{Wielgus:2022heh,Yfantis:2023wsp}, we additionally fix the observer inclination to $\theta_{\rm obs}=20^\circ$ in this initial investigation of an inspiraling hotspot.

\subsection{Fitting the Primary Loop with an Inspiraling Hotspot}\label{sec:K1results}

We start by validating our methodology on a single segment ($K=1$) over the same time interval as analyzed in \citet{Yfantis:2023wsp}, up to $t_{\rm obs}\sim230\,M$. In \citet{Yfantis:2023wsp}, the preference between a Keplerian and sub-Keplerian orbit is found to depend on the assumed magnetic field polarity, a distinction our model cannot make since Stokes $Q$ and $U$ are invariant under $\vec B \to - \vec B$.\footnote{The lack of this invariance in \citet{Yfantis:2023wsp} is a consequence of plasma effects and the internal Faraday depth. Here we have assumed a negligible Faraday depth in the emitting zone (see, for instance,~\cite{2023ApJ...950...38E}).} In our analysis, reported in the upper block of Table~\ref{tab:fit_results}, we find results consistent with their default-polarity variants. Models are labeled by variant: $A$ sets $\beta_r=\beta_\phi=1$; $B$ frees $\beta_\phi$; $C$ frees $\beta_r$; $D$ frees both; $E$ frees both with one $\alpha$ shared across segments; $F$ frees $\beta_r$ with shared $\alpha$; and $G$ also shares a background offset $(Q_{\rm sha},U_{\rm sha})$ across segments. The subscript indicates $K$; a prime denotes the shorter $t_{\rm obs}\sim230\,M$ subset.

Allowing for deviations from Keplerian motion, the model in which only $\beta_\phi$ is free ($B_1'$) is strongly disfavored by the Bayesian evidence relative to the Keplerian model ($A_1'$), although its fit is marginally better. Using the same observation window, we now broaden the analysis to include inspiraling trajectories through two additional variants, $C_1'$, where $\beta_r$ is free, and $D_1'$, where both Keplerianity parameters are free. Model $C_1'$ has the greatest evidence overall and the best $\chi^2_{\rm eff}$ of the $K=1$ models. Whenever the azimuthal parameter $\beta_\phi$ is left free, the evidence penalizes the run because of a degeneracy between the sampled initial radius and $\beta_\phi$, visible in the $K=2$ posterior of Fig.~\ref{fig:K2_corner_nonkep}. Freeing the radial parameter $\beta_r$ alongside it weakens this degeneracy but does not remove it, so the fully non-Keplerian model $D_1'$ is still disfavored relative to $A_1'$ although it produces a better fit. 

\begin{table*}[]
    \centering
    \caption{Fit results for the two subsets of the 2017 April 11 ALMA
    observation, both sampled at $80\,$s cadence with fixed $a=0.87$ and
    $\theta_{\rm obs}=20^\circ$. Rows are labeled by variant, where $A$ has
    $\beta_r=\beta_\phi=1$, $B$ frees $\beta_\phi$, $C$ frees $\beta_r$, $D$
    frees both, $E$ frees both with a single $\alpha$ shared across
    segments, $F$ frees $\beta_r$ with a single shared $\alpha$, and $G$
    additionally shares a single background offset $(Q_{\rm sha},U_{\rm sha})$
    across segments. The subscript gives $K$, and a prime marks the shorter
    $t_{\rm obs}\sim230\,M$ subset. The $K=1$ runs use a $12$-temperature geometric ladder with $64$ walkers, and the $K=2$ and $K=3$ runs the $45$-temperature densified ladder, with $64$ walkers on the $t_{\rm obs}\sim350\,M$ subset and $256$ on the $t_{\rm obs}\sim230\,M$ subset. At each model's highest-posterior sample from the cold chain we evaluate $\chi^2_{\rm eff}$, and the ESS is the minimum full-chain effective sample size across all sampled dimensions, i.e. the cold chain is thinned by the greatest autocorrelation time among all the parameters. Within the $t_{\rm obs}\sim230\,M$ subset the data marginally prefer the mild inspiral $C_1'$ over the Keplerian case $A_1'$, by $\Delta\ln Z\approx1.3$, and disfavor a second kinematic segment, with the best two-segment variant $G_2'$ falling $\Delta\ln Z\approx17$ below $C_1'$. Within the $t_{\rm obs}\sim350\,M$ subset the single-segment fit $A_1$ is strongly rejected, the best two-segment fit is produced by $D_2$, and the highest evidence by the simpler model $E_2$. The data do not prefer a three-kinematic-segment model ($D_3$, $E_3$, and $H_3$) over any model with two segments. Evidences are comparable only within a subset, since the two are computed from different data and are
    therefore not on a common scale.}
    \label{tab:fit_results}
    \setlength{\tabcolsep}{5pt}
    \renewcommand{\arraystretch}{1.3}
    \begin{tabular}{llcccccccc}
        \toprule
        Label & Variant & $K$ & $N_{\rm dim}$ & $\ln\mathcal{L}_{\rm best}$ & $\chi^2$ & $\chi^2_{\rm eff}$ & $\ln Z_{\rm ti} \pm \sigma$ & $\ln Z_{\rm ss} \pm \sigma$ & ESS\\
        \midrule
        \multicolumn{10}{l}{\textit{$t_{\rm obs}\sim230\,M$ subset, $N=44$ $(n_d=88)$}}\\
        \addlinespace[2pt]
        $A_1'$ & $\beta_r=\beta_\phi=1$            & $1$ &  $9$ & $278.999$ & $34.741$ & $0.440$ & $246.471 \pm 3.114$ & $246.815 \pm 0.035$ & $37\,937$ \\
        $B_1'$ & $\beta_r=1$, $\beta_\phi$ free    & $1$ & $10$ & $279.304$ & $34.157$ & $0.438$ & $227.567 \pm 4.241$ & $228.447 \pm 0.241$ & $34\,559$ \\
        $C_1'$ & $\beta_r$ free, $\beta_\phi=1$    & $1$ & $10$ & $281.742$ & $29.384$ & $0.377$ & $247.804 \pm 3.422$ & $248.117 \pm 0.034$ & $35\,681$ \\
        $D_1'$ & $\beta_r$, $\beta_\phi$ free      & $1$ & $11$ & $281.780$ & $29.363$ & $0.381$ & $244.101 \pm 4.032$ & $244.467 \pm 0.041$ & $31\,981$ \\
        \addlinespace[3pt]
        $D_2'$ & $\beta_r$, $\beta_\phi$ free      & $2$ & $17$ & $286.407$ & $19.788$ & $0.279$ & $222.082 \pm 0.801$ & $222.170 \pm 0.654$ & $121\,751$ \\
        $F_2'$ & $\beta_r$ free, $\beta_\phi=1$ (global $\alpha$) & $2$ & $14$ & $284.013$ & $24.771$ & $0.335$ & $222.237 \pm 0.784$ & $222.312 \pm 0.593$ & $145\,830$ \\
        $G_2'$ & $\beta_r$ free, $\beta_\phi=1$ (global $\alpha$ and $Q_{\rm sha},U_{\rm sha}$) & $2$ & $12$ & $283.487$ & $26.082$ & $0.343$ & $230.999 \pm 0.596$ & $231.016 \pm 0.584$ & $131\,655$ \\
        \midrule
        \multicolumn{10}{l}{\textit{$t_{\rm obs}\sim350\,M$ subset, $N=62$ $(n_d=124)$}}\\
        \addlinespace[2pt]
        $A_1$ & $\beta_r=\beta_\phi=1$             & $1$ &  $9$ & $106.907$ & $616.720$ & $5.363$ & $73.842 \pm 0.774$  & $73.901 \pm 0.022$  & $156\,371$ \\
        $A_2$ & $\beta_r=\beta_\phi=1$             & $2$ & $13$ & $300.592$ & $231.351$ & $2.084$ & $251.172 \pm 0.778$ & $251.265 \pm 0.042$ & $176\,545$ \\
        $B_2$ & $\beta_r=1$, $\beta_\phi$ free     & $2$ & $15$ & $361.578$ & $109.787$ & $1.007$ & $264.885 \pm 0.874$ & $264.919 \pm 0.691$ & $5\,367$   \\
        $C_2$ & $\beta_r$ free, $\beta_\phi=1$     & $2$ & $15$ & $378.146$ &  $76.506$ & $0.702$ & $277.101 \pm 0.767$ & $277.140 \pm 0.474$ & $6\,483$   \\
        $D_2$ & $\beta_r$, $\beta_\phi$ free       & $2$ & $17$ & $385.247$ &  $62.130$ & $0.581$ & $296.798 \pm 0.863$ & $296.885 \pm 0.363$ & $159\,735$ \\
        $E_2$ & $\beta_r$, $\beta_\phi$ free (global $\alpha$) & $2$ & $16$ & $381.636$ & $69.802$ & $0.646$ & $297.375 \pm 0.874$ & $297.469 \pm 0.426$ & $165\,737$ \\
        $D_3$ & $\beta_r$, $\beta_\phi$ free       & $3$ & $23$ & $393.670$ &  $45.076$ & $0.446$ & $233.125 \pm 1.580$ & $233.024 \pm 1.793$ & $193\,443$ \\
        $E_3$ & $\beta_r$, $\beta_\phi$ free (global $\alpha$) & $3$ & $21$ & $387.043$ & $58.409$ & $0.567$ & $233.021 \pm 0.922$ & $232.976 \pm 1.654$ & $6\,254$ \\
        $H_3$ & $\beta_r$, $\beta_\phi$ free (global $\alpha$ and $Q_{\rm sha},U_{\rm sha}$) & $3$ & $17$ & $368.768$ & $94.759$ & $0.886$ & $280.359 \pm 0.672$ & $280.368 \pm 0.824$ & $51\,498$ \\
        \bottomrule
    \end{tabular}
\end{table*}

These $K=1$ runs use a $12$-temperature geometric ladder, following the default \texttt{reddemcee} temperature configuration, all assuming the priors of Table~\ref{tab:priors} and converging within $40\,000$ steps. For the runs of Sec.~\ref{sec:K2_results}, however, the models that free $\beta_\phi$ develop multi-modal posteriors, due to the degeneracy of this parameter with the initial position, and require a denser ladder. We therefore use a $45$-temperature ladder for all runs in that section, with a higher density of temperatures in the range where the swap acceptance fraction drops and a phase transition occurs. This denser ladder is what allows those runs to converge in a comparable number of steps, and we refer to it as the ``densified'' ladder.

To test whether a single segment suffices on the $t_{\rm obs}\sim230\,M$ dataset, we run three two-segment analyses on it with the densified ladder. The first repeats the $D$ variant with two segments ($D_2'$). Since the evidence on this dataset consistently favors models with $\beta_\phi$ held at its Keplerian value, as seen for the single-segment runs above, we also run two two-segment versions of the kinematics of $C_1'$, the highest-evidence model. One ($F_2'$) shares a single $\alpha_\nu$ across both segments and the other ($G_2'$) shares both $\alpha_\nu$ and the background offset $(Q_{\rm sha},U_{\rm sha})$. All three fall well below $C_1'$ in evidence, the best of them by $\ln Z(C_1')-\ln Z(G_2')\approx17$, confirming that the data within $t_{\rm obs}\sim230\,M$ do not require two segments.

The single-segment $(K=1)$ analyses, conducted over the same observation interval as in \citet{Yfantis:2023wsp}, yield mild evidence in favor of motion that includes a radial inflow component. Nevertheless, as indicated by the corresponding light-colored curves in Fig.~\ref{fig:K1vsK2}, continuing the best-fitting single-segment models beyond the dashed vertical line results in a marked degradation of the fit. Indeed, across the full dataset shown in Fig.~\ref{fig:K1vsK2}, no single-segment model provides an adequate fit over the entire time span. In the next section we extend the analysis beyond the primary loop, fitting the $t_{\rm obs}\sim350\,M$ subsample, and allowing the kinematics to change once along the trajectory ($K=2$).

\subsection{Fitting the Evolving $Q$--$U$ Loop: Its Inner Structure}\label{sec:K2_results}

On the $t_{\rm obs}\sim350\,M$ subsample we compare the nine variants in the lower block of Table~\ref{tab:fit_results}, all run with the priors of Table~\ref{tab:priors} and the densified ladder. Variant $A_1$ is the single-segment Keplerian model of the previous section, now fitted to the longer window. The remaining five use two kinematic segments and differ in which Keplerianity parameters are freed. The variant $A_2$ keeps both fixed at their Keplerian values, $B_2$ frees the angular parameter $\beta_\phi$, $C_2$ frees the radial parameter $\beta_r$, and $D_2$ frees both, each of these sampling one spectral index per segment. The variant $E_2$ repeats $D_2$ with a single spectral index shared across both segments, which removes one dimension from the fit, and $D_3$ extends $D_2$ to three segments. Splitting the Keplerian trajectory into two segments ($A_2$) substantially improves on $A_1$ but still falls $\Delta\ln Z\approx46$ below $E_2$. The two-segment counterpart of the preliminary best model, $C_2$, also underperforms, by $\ln Z(E_2)-\ln Z(C_2)\approx20$ and in the quality of the fit. The fully non-Keplerian cases are preferred by the data, with the simpler variant $E_2$ having the highest evidence. The difference between $E_2$ and $D_2$, $\Delta\ln Z\approx0.6$, is not statistically significant, which is enough to conclude that the extra spectral index in $D_2$ is not required, so a single spectral index through the whole subsample suffices, as in $E_2$. 

Given the model’s flexibility, we also evaluated an extension to a three-segment configuration ($D_3$, $E_3$, and $H_3$). However, for the present dataset, there is no statistical support for additional segments. In particular, $D_3$, the most general model, achieves the highest raw log-likelihood, but its Bayesian evidence lies $\Delta\ln Z\approx64$ below $E_2$. Its posterior collapses toward the two-segment solution, with the first break unconstrained near the start of the track and the second at the $K=2$ break time, so the additional segment enlarges the prior volume more than it improves the fit. The same conclusion holds for $E_3$ and $H_3$, indicating that the $t_{\rm obs}\sim 350\,M$ data do not require more than two segments.

\begin{figure*}[]
    \centering
    \includegraphics[width=\textwidth]{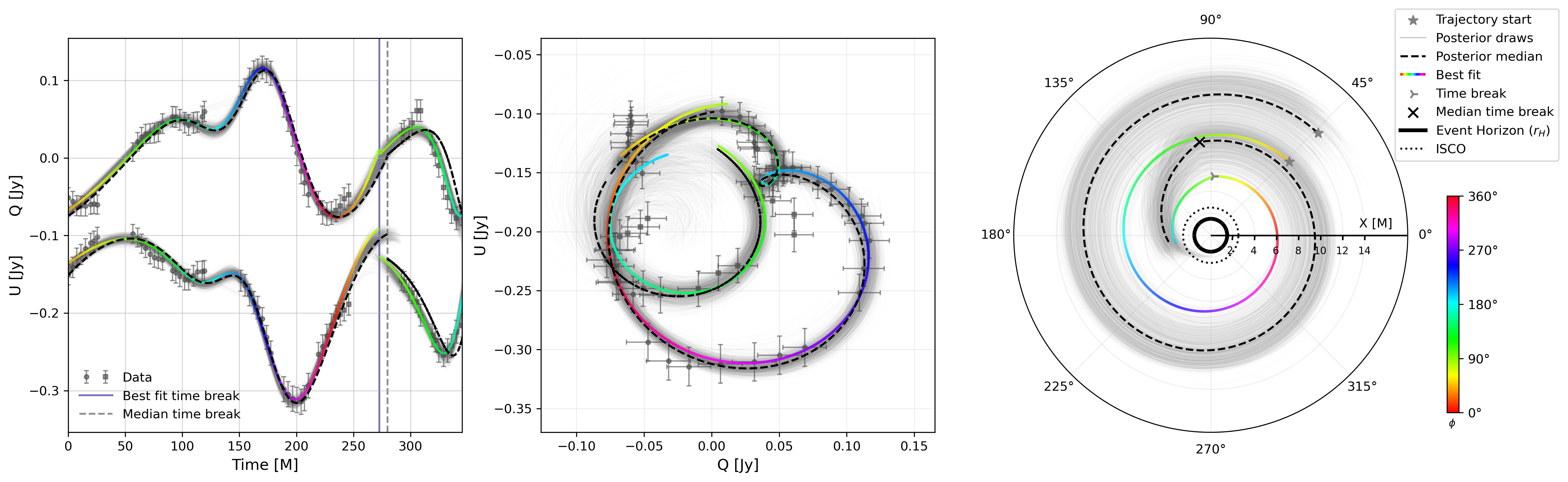}
    \caption{Non-Keplerian $E_2$ model results showing the posterior bands for $Q(t)$ and $U(t)$ (left), the $Q$--$U$ fit (center), and the corresponding trajectory of the hotspot in the equatorial plane (right), with the radial coordinate in units of $M$ and azimuth measured from the $X$ axis. The data are shown in gray, with errors calculated using Eq.~\ref{eq:error}, drawn as circles within $t_{\rm obs}\sim230\,M$ and as squares beyond it. The best-fit curve is colored according to the instantaneous azimuthal position of the hotspot $\phi$, shared across all three panels, and the median reconstructed curve is shown as a black dashed line. In the left panel the violet solid and gray dashed vertical lines mark the best-fit and median break times, respectively. To visualize the posterior distribution in the $Q$--$U$ plane, we show $1\,000$ random samples from the posterior in light gray. In the right panel the dashed black line shows the inspiral plotted from the median of the distribution, while the colored line shows the best-fit trajectory, with the trajectory start marked by $\bigstar$ and the best-fit and median break times by \protect\triright{} and $\boldsymbol{\times}$, respectively. The event horizon and the ISCO are drawn as a solid and a dotted circle, respectively, at their values for $a=0.87$. The light-gray lines represent the trajectories from the same $1\,000$ random samples from the posterior. The observed emission does not tightly constrain the path the hotspot takes, since those same samples give $Q$--$U$ values very close to the median in the left and center panels.}
    \label{fig:K2_QU_traj_nonkep}
\end{figure*}

\begin{figure*}[]
    \centering
    \includegraphics[width=\textwidth]{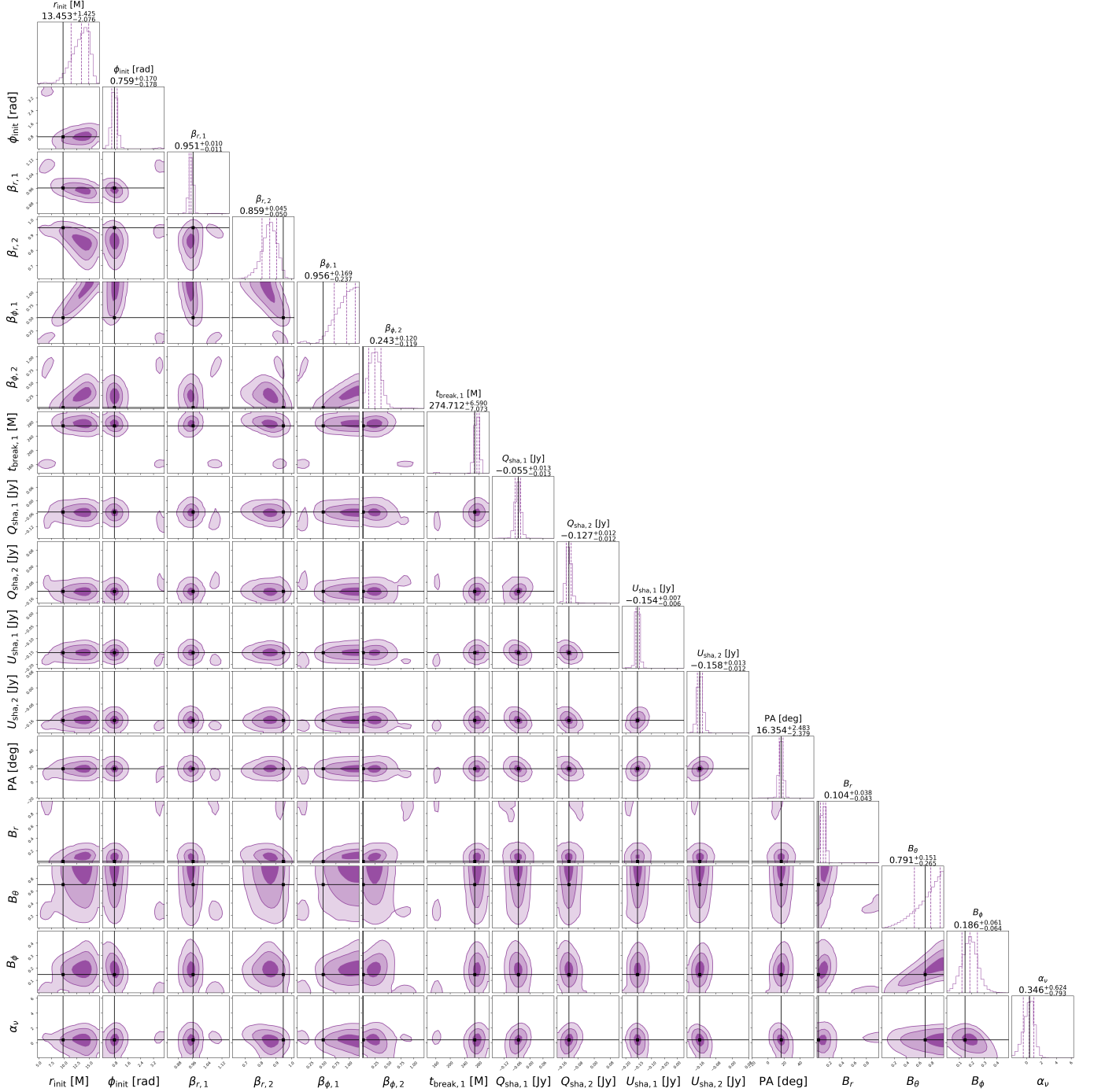}
    \caption{Posterior corner plot for the non-Keplerian $E_2$ model. Black lines mark the best-fit parameter values. The spectral index is calculated as $\alpha_\nu = 2\alpha - 3$, and the break time is computed from the time variables $\tau_i$ using Eqs.~\ref{eq:time_weights}--\ref{eq:fixedK_duration}.}
    \label{fig:K2_corner_nonkep}
\end{figure*}

The preference for a single spectral index suggests that the emission's spectral shape does not change appreciably over the observation. The resulting $Q(t),U(t)$ plotted values for the $E_2$ variant of the model are shown in Fig.~\ref{fig:K2_QU_traj_nonkep}, along with the corresponding corner plot in Fig.~\ref{fig:K2_corner_nonkep}. Among the $K=2$ models in the lower block of Table~\ref{tab:fit_results}, the best $\chi^2$ and $\chi^2_{\rm eff}$ correspond to $D_2$, but the evidence does not support this extra freedom, as explained above. All distributions from the runs for this extended dataset are thinned by the largest autocorrelation time among the parameters within each case. The results before and after thinning are consistent in their distribution characteristics and calculated evidence, indicating that the invariant distribution of the posterior has been reached.

The posterior distributions for model $E_2$, shown in Fig.~\ref{fig:K2_corner_nonkep}, describe how the flow changes across the two segments. Values quoted below are posterior medians with $16$th and $84$th percentiles. Because the Keplerian radial velocity vanishes outside the ISCO, $\beta_r<1$ makes the inflow a fraction $(1-\beta_r)$ of the free-fall value at the same radius. We infer $\beta_{r,1}=0.951^{+0.010}_{-0.011}$ and $\beta_{r,2}=0.859^{+0.045}_{-0.050}$, an inward drift of about $5\%$ of the local free-fall speed before the break and $14\%$ after it. Both the best-fit and median tracks terminate well outside the ISCO, as shown in the right panel of Fig.~\ref{fig:K2_QU_traj_nonkep}.

The angular component behaves differently. We find $\beta_{\phi,1}=0.96^{+0.17}_{-0.24}$, consistent with the Keplerian motion assumed in previous fits of the primary loop~\citep{Wielgus:2022heh,Yfantis:2023wsp}, and $\beta_{\phi,2}=0.24^{+0.12}_{-0.12}$. In the first segment the angular fraction is constrained an order of magnitude less tightly than the radial one, with the gap narrowing in the second. Sub-Keplerian rotation accompanied by a radial inflow is a generic feature of hot, advection-dominated accretion flows~\citep{Narayan:1994xi}, and is also found for orbiting flux tubes in GRMHD simulations of magnetically arrested disks~\citep{Porth:2020txf}.

Beyond the kinematics, the per-segment background offsets recover the static background polarization reported independently in the literature. For segment $1$ of model $E_2$ the sampled offsets are $Q_{{\rm sha},1}=-0.055^{+0.013}_{-0.013}$ Jy and $U_{{\rm sha},1}=-0.154^{+0.007}_{-0.006}$ Jy, giving a magnitude $|\mathcal{P}_{{\rm sha},1}|=0.164^{+0.003}_{-0.003}$ Jy. This is within $0.005$\,Jy of the $|\mathcal{P}_{\rm sha}|=0.16$\,Jy static component measured by \citet{Wielgus:2022heh}, the same value that \citet{Levis:2023tpb} subtract from this dataset before fitting. Nothing in our setup imposes it, since $(Q_{{\rm sha},i},U_{{\rm sha},i})$ are free parameters with a $[-1,1]$ box prior, so the agreement is an emergent property of the fit and corroborates the amplitude calibration described above.

The angle of the same component requires the rotation of Eq.~\ref{eq:evparot} before it can be compared. Because the background offset is applied before the global rotation, the sampled $(Q_{{\rm sha},i},U_{{\rm sha},i})$ live in the emission frame, which is fixed to the disk rather than to the observer. The corresponding position angle is $\chi_{{\rm sha},1}^{\rm model}=\tfrac12\arg(Q_{{\rm sha},1} + \mathrm{i} U_{{\rm sha},1})=-54.89^{+2.37}_{-2.46}$ degrees, which is not the observable quantity. Rotating by the inferred $\text{PA}=16.35^{+2.48}_{-2.38}$ degrees, comparable to the rotation angles inferred in previous Bayesian analyses of this dataset~\citep{Yfantis:2023wsp}, gives the sky-frame value $\chi_{{\rm sha},1}^{\rm sky}=-38.50^{+0.74}_{-0.73}$ degrees, which is what should be set against the $-37$ degrees reported previously~\citep{Wielgus:2022heh}. The comparison is only approximate, since a time-variable rotation measure has been reported at millimeter wavelengths in Sgr~A* during the 2017 April 11 flaring period~\citep{Wielgus:2022heh,Wielgus:2023akf}, which our single non-varying PA cannot represent.

The spectral index is the least constrained parameter of the fit. For $E_2$ we obtain $\alpha=1.67^{+0.31}_{-0.40}$, or $\alpha_\nu=0.35^{+0.62}_{-0.79}$ in the convention of Sec.~\ref{sec:observed_polarization}, which is consistent with the value measured for Sgr~A* at millimeter wavelengths immediately after the flare~\citep{EventHorizonTelescope:2022ago}, $\alpha_\nu=0.25\pm0.10$ in the convention we use in this work. The best $K=2$ fit comes from $D_2$, which samples one index per segment, but its indices are no better constrained and the evidence does not require the extra freedom, so we cannot say whether the emitting electron population evolves across the break. We leave that study to future work. The results from the $D_2$ model can be found in Appendix~\ref{sec:K2_more_results} in Fig.~\ref{fig:K2_qu_nonkep_general}, along with its posterior distributions shown in Fig.~\ref{fig:K2_corner_nonkep_general}.

\section{Conclusions}\label{sec:conclusions}

We analyzed the 2017 April 11 ALMA $229$~GHz observation of Sgr~A*~\citep{Wielgus:2022heh}, which captured a flaring event followed by a period of variability in the source's polarized flux. In a previous study of this polarized emission, models of a hotspot on a circular orbit were used to interpret the observations~\citep{Yfantis:2023wsp}. Using the hotspot model from \citet{Ruales:2026xjb}, which allows for a non-zero radial velocity component, we find evidence that the data prefer an inspiraling trajectory.

When studying the same time range as in \citet{Yfantis:2023wsp}, we find that the primary loop mildly prefers an inspiral with a small radial velocity component. However, such a mild inspiral cannot explain the rest of the data when computed beyond the primary loop. The posterior median of the two-segment model $E_2$ places the change in kinematics at $t_{\rm break,1}=274.7^{+6.6}_{-7.1}\,M$ in coordinate time of emission, or $\approx280\,M$ as seen by the observer, beyond the $t_{\rm obs}\sim230\,M$ covered by previous analyses, so a fit restricted to that window cannot resolve it. These findings suggest that the data prefer a trajectory with an increasing radial velocity, as can be deduced from the path shown in the right panel of Fig.~\ref{fig:K2_QU_traj_nonkep}. The agreement with previous studies in this first portion of the data occurs because of a degeneracy between models when the hotspot is far from the ISCO. Different hotspot trajectories within the posterior distribution produce similar polarization signatures (see the posterior samples in Figs.~\ref{fig:K2_QU_traj_nonkep} and~\ref{fig:K2_qu_nonkep_general}). However, as the hotspot gets closer to the ISCO, these degeneracies start to break, and if only one segment is used, the extended dataset cannot be fitted well (see Fig.~\ref{fig:K1vsK2}). 

Using the multi-segmented inspiraling model, we find that the highest-evidence variant is a fully non-Keplerian ($\beta_r$, $\beta_\phi$ free) trajectory with a single $\alpha_\nu$ shared across segments (model $E_2$), favored over its two-segment Keplerian counterpart by $\Delta\ln Z\approx46$. With it we successfully fit up to $t_{\rm obs}\sim350\,M$, roughly $120\,M$ or $0.7$\,hr beyond the range covered by previous studies. Across the break, the flow also loses most of its rotational support, with $\beta_\phi$ falling from $\approx1.0$ to $\approx0.2$, though the inferred track still terminates outside the ISCO. This kinematic jump is also found in model $D_2$, shown in Fig.~\ref{fig:K2_corner_nonkep_general}. This can be understood as a bright feature being carried with the advection-dominated accretion flow, accelerating toward the event horizon~\citep{Narayan:1994xi} rather than being expelled from the system through magnetic buoyancy~\citep{Porth:2020txf}. However, a strictly equatorial model may not be suitable to capture the latter effect. The obtained spectral index distribution is consistent with the value measured immediately after the 2017 April 11 X-ray flare, which in our convention corresponds to $\alpha_\nu=0.25\pm0.10$~\citep{EventHorizonTelescope:2022ago}. Moreover, our fitting procedure allowed us to estimate the magnetic field configuration without strong priors. We confirm the dominance of the vertical magnetic field component that was argued for in \citet{Abuter:2018uum} and \citet{Wielgus:2022heh}, while other modeling attempts typically adopt it as an assumption~\citep{Yfantis:2023wsp,Levis:2023tpb,Tlemissov:2026kuu}.

This first attempt to fit the inner loop structure and the changing morphology of the $Q$--$U$ loops motivates further theoretical work. In particular, a model that accelerates smoothly rather than in segments would be a natural next step, as would sampling the spin and the observer's inclination, which we held fixed at values taken from previous fits of this dataset~\citep{Wielgus:2022heh,Yfantis:2023wsp}, since the inferred kinematics are conditional on that geometry. Furthermore, the global PA rotation is an approximation to a highly variable parameter related to the rotation measure~\citep{Wielgus:2022heh,Wielgus:2023akf}. The model cannot simulate Faraday rotation directly, but a single global PA appears to absorb it well enough to place the loop center correctly, as can be seen in the $Q$--$U$ fits in Figs.~\ref{fig:K2_QU_traj_nonkep} and~\ref{fig:K2_qu_nonkep_general}. Physically, this corresponds to modeling only the static external component of the Faraday screen~\citep{Wielgus:2023akf}. The inferred angle is comparable to the rotation angles obtained in previous Bayesian analyses of this dataset~\citep{Yfantis:2023wsp}.

In our model the emitting hotspot does not evolve in its comoving frame. Thus, effects such as gravitational redshift and Doppler boost away from the line of sight are responsible for the difference in the appearance of the first and the second $Q$--$U$ loop in the dataset. Alternatively, these differences could be attributed to internal depolarization as a consequence of plasma instabilities and shearing in a differentially rotating flow, or to an overall change of the emitted flux with radiative cooling. There is some theoretical support for hotspots related to magnetic flux tubes surviving for up to two orbits without much internal change~\citep{Porth:2020txf} and the cooling timescale may be sufficiently long at millimeter wavelengths~\citep{Yfantis:2023wsp}. Nonetheless, clarifying this degeneracy will require simultaneous observations across a wide range of frequencies or actually resolving the orbit of the hotspot~\citep{Johnson:2023ynn,Emami:2022ydq}.

Kinematics from general relativistic magnetohydrodynamic simulations~\citep{Porth:2020txf,Ripperda:2021zpn,Najafi-Ziyazi:2023oil} or including variability through a stochastic model~\citep{Cardenas-Avendano:2026bow} could guide modeling and provide informed priors for such analyses, increasing the efficiency of the method. Models of this kind will be needed to meet the demands of interpreting future high angular resolution observations, such as those of the next-generation Event Horizon Telescope~\citep[ngEHT;][]{Johnson:2023ynn,Emami:2022ydq} and the Black Hole Explorer~\citep[BHEX;][]{Johnson:2024ttr,Lupsasca:2024xhq}. By developing improved theoretical models of hotspot kinematics and emission and testing them against future observations, we will be able to better understand both accretion processes and the effects of strong gravity in the Galactic Center.

\begin{acknowledgments}
We thank A.~Yfantis and L.~Keeble for valuable comments. MW is supported by a Ramón y Cajal grant RYC2023-042988-I from the Spanish Ministry of Science and Innovation and acknowledges financial support from the Severo Ochoa grant CEX2021-001131-S funded by MCIN/AEI/ 10.13039/501100011033. 
DG was supported by the Gordon and Betty Moore Foundation (Grant \#12987) and the National Science Foundation (AST-2307887), and by the Black Hole Initiative through the Gordon and Betty Moore Foundation (Grant \#13526) and the John Templeton Foundation (Grant \#63445). The opinions expressed in this publication are those of the authors and do not necessarily reflect the views of these Foundations and organizations. 
Claude Fable~$5$ was used to suggest language improvements within the manuscript and to optimize the performance of existing analysis code. The output was verified to be numerically identical to the original human-written code. The authors take full responsibility for all results presented in this work.
Computations were performed using the Wake Forest University (WFU) High Performance Computing Facility, a centrally managed computational resource available to WFU researchers including faculty, staff, students, and collaborators~\citep{WakeHPC}.
\end{acknowledgments}

\appendix

\section{Lensing Magnification of the Hotspot}
\label{sec:magnification}

The hotspot in the model presented in \citet{Ruales:2026xjb} is understood as a point source on the equatorial plane of the disk, with Boyer--Lindquist coordinates $(r,\phi)$. The mapping of the observer's screen $(\alpha,\beta)$ onto the disk is a transformation which is geometry-dependent. To capture the emission intensity profile of a point source, we multiply the computed polarization by a magnification factor calculated from a ratio of areas as done by \citet{Zhu:2026jom}. The \texttt{AART} method involves constructing a grid for each lensing band $n$, in which every screen pixel $(\alpha,\beta)$ has a corresponding Cartesian $(x,y)$ coordinate (or $(r,\phi)$) in the physical disk plane. This mapping has a Jacobian
\begin{equation}\label{eq:jacobian}
    J\equiv\frac{\partial(x,y)}{\partial(\alpha,\beta)}.
\end{equation}
A discrete screen pixel $d\alpha\,d\beta$ corresponds to a flat-coordinate disk region $dA_{\rm flat}=dx\,dy=\ab{\det J}\,d\alpha\,d\beta$, leading to the definition of a ``local'' magnification factor
\begin{equation}\label{eq:mulocal}
    \mu_{\rm local}(\alpha,\beta)\equiv\frac{dA_{\rm screen}}{dA_{\rm flat}}=\frac{1}{\ab{\det J}},
\end{equation}
which depends only on the geometry and not on any property of the hotspot. Equation~\ref{eq:mulocal} is a good approximation for regions far outside the ISCO. However, for an inspiraling hotspot that potentially reaches the horizon, we need a ``proper'' geometrical correction. The equatorial $(\theta=\pi/2)$ Boyer--Lindquist metric gives the proper area element
\begin{equation}
    dA_{\rm proper}=\sqrt{g_{rr}g_{\phi\phi}}\;dr\,d\phi=\sqrt{\frac{A(r)}{\Delta(r)}}\;dr\,d\phi,
\end{equation}
with $\Delta(r)=r^2-2r+a^2$ and $A(r)=(r^2+a^2)^2-a^2\Delta(r)$. Once again, the ratio of the two areas defines a proper area factor
\begin{equation}\label{eq:propareafactor}
    f(r)\equiv\frac{dA_{\rm proper}}{dA_{\rm flat}}=\frac{\sqrt{A(r)/\Delta(r)}}{r}.
\end{equation}
Combining Eqs.~\ref{eq:mulocal} and~\ref{eq:propareafactor}, the magnification of a point source at disk position $(r,\phi)$ and imaged at screen position $(\alpha,\beta)$, is
\begin{equation}\label{eq:muproper}
    \mu_{\rm proper}(r,\phi)=\frac{dA_{\rm screen}}{dA_{\rm proper}}=\frac{\mu_{\rm local}(\alpha,\beta)}{f(r)},
\end{equation}
which is analytic in $r$ and is evaluated on the fly for every point of the hotspot's trajectory. The computed linear polarization is scaled by this factor as
\begin{equation}
    Q(t)\rightarrow\mu_{\rm proper}\big(r(t)\big)\,Q(t),\,
    U(t)\rightarrow\mu_{\rm proper}\big(r(t)\big)\,U(t).
\end{equation}
This process is necessary because \texttt{AART} inherently assumes an area for the point source when it creates the screen grid before the backward ray-tracing happens~\citep{Cardenas-Avendano:2022csp}. Thus, $\mu_{\rm proper}$ rescales the polarized emission profile to a true point source.

\section{Best-Fit $K=2$ Model}
\label{sec:K2_more_results}

\begin{figure*}[h!]
    \centering
        \includegraphics[width=\textwidth]{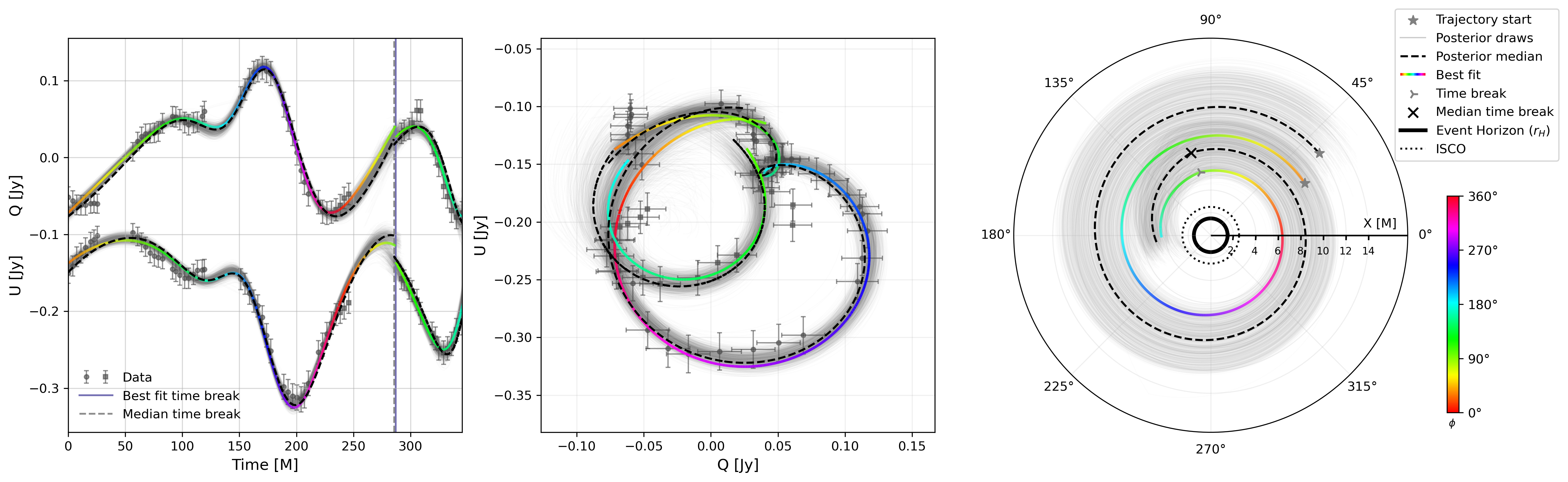}%
            \caption{Non-Keplerian $D_2$ model results showing the posterior bands for $Q(t)$ and $U(t)$ (left), the $Q$--$U$ fit (center), and the corresponding trajectory of the hotspot in the equatorial plane (right), with the radial coordinate in units of $M$ and azimuth measured from the $X$ axis. The data are shown in gray, with errors calculated using Eq.~\ref{eq:error}, drawn as circles within $t_{\rm obs}\sim230\,M$ and as squares beyond it. The best-fit curve is colored according to the instantaneous azimuthal position of the hotspot $\phi$, shared across all three panels, and the median reconstructed curve is shown as a black dashed line. In the left panel the violet solid and gray dashed vertical lines mark the best-fit and median break times, respectively. To visualize the posterior distribution in the $Q$--$U$ plane, we show $1\,000$ random samples from the posterior in light gray. In the right panel the dashed black line shows the inspiral plotted from the median of the distribution, while the colored line shows the best-fit trajectory, with the trajectory start marked by $\bigstar$ and the best-fit and median break times by \protect\triright{} and $\boldsymbol{\times}$, respectively. The event horizon and the ISCO are drawn as a solid and a dotted circle, respectively, at their values for $a=0.87$. The light-gray lines represent the trajectories from the same $1\,000$ random samples from the posterior. The observed emission does not tightly constrain the path the hotspot takes, since those same samples give $Q$--$U$ values very close to the median in the left and center panels.}
    \label{fig:K2_qu_nonkep_general}
\end{figure*}

\begin{figure*}[]
    \centering        \includegraphics[width=\textwidth]{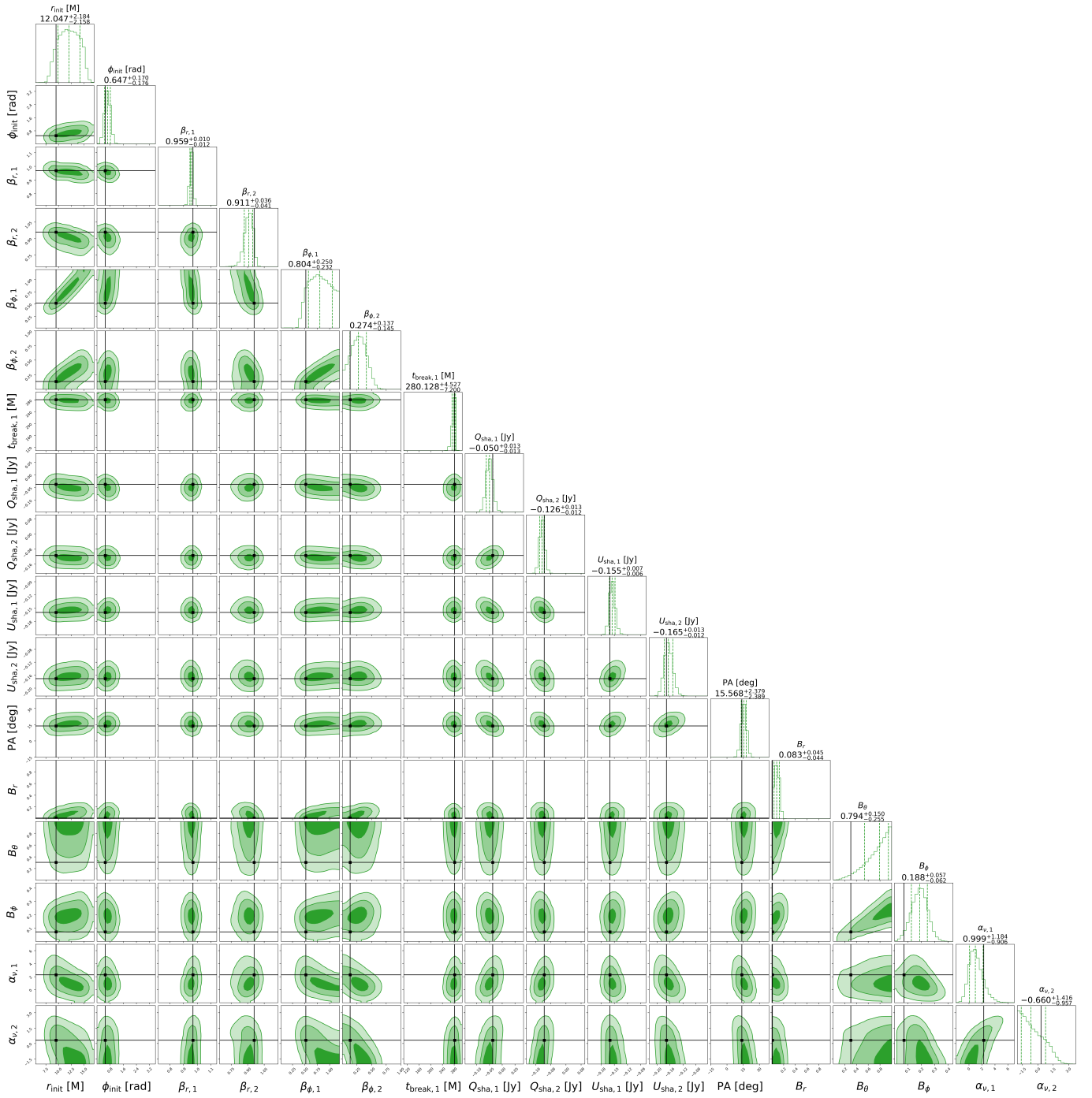}
            \caption{Posterior corner plot for the non-Keplerian $D_2$ model. Black lines mark the best-fit parameter values. The spectral index is calculated as $\alpha_\nu = 2\alpha - 3$, and the break time is computed from the time variables $\tau_i$ using Eqs.~\ref{eq:time_weights}--\ref{eq:fixedK_duration}.}
    \label{fig:K2_corner_nonkep_general}
\end{figure*}

This appendix contains the results of the $D_2$ model, the median and best-fit plots in Fig.~\ref{fig:K2_qu_nonkep_general} and its posterior distribution corner plot in Fig.~\ref{fig:K2_corner_nonkep_general}.

\bibliography{refs}{}

@article{Zhu:2026jom,
    author = "Zhu, Qing-Hua",
    title = "{Flux enhancement of corotating hot spots in accretion disks from the slow-light effect}",
    eprint = "2604.09046",
    archivePrefix = "arXiv",
    primaryClass = "gr-qc",
    doi = "10.1103/hgsl-lnls",
    journal = {\prd},
    volume = "114",
    number = "4",
    pages = "044057",
    year = "2026"
}

@article{Taubin1991,
    author = "Taubin, Gabriel",
    title = "{Estimation of planar curves, surfaces, and nonplanar space curves defined by implicit equations with applications to edge and range image segmentation}",
    doi = "10.1109/34.103273",
    journal = {ITPAM},
    volume = "13",
    number = "11",
    pages = "1115--1138",
    year = "1991"
}

@article{Vincent:2022fwj,
    author = "Vincent, Frederic H. and Gralla, Samuel E. and Lupsasca, Alexandru and Wielgus, Maciek",
    title = "{Images and photon ring signatures of thick disks around black holes}",
    eprint = "2206.12066",
    archivePrefix = "arXiv",
    primaryClass = "astro-ph.HE",
    doi = "10.1051/0004-6361/202244339",
    journal = {\aap},
    volume = "667",
    pages = "A170",
    year = "2022"
}

@article{Bardeen:1972fi,
    author = "Bardeen, James M. and Press, William H. and Teukolsky, Saul A",
    title = "{Rotating black holes: Locally nonrotating frames, energy extraction, and scalar synchrotron radiation}",
    reportNumber = "OAP-288",
    doi = "10.1086/151796",
    journal = {\apj},
    volume = "178",
    pages = "347",
    year = "1972"
}

@article{Cunningham:1975zz,
    author = "Cunningham, C. T.",
    title = "{The effects of redshifts and focusing on the spectrum of an accretion disk around a Kerr black hole}",
    doi = "10.1086/154033",
    journal = {\apj},
    volume = "202",
    pages = "788--802",
    year = "1975"
}

@article{Gelles:2021kti,
    author = "Gelles, Zachary and Himwich, Elizabeth and Palumbo, Daniel C. M. and Johnson, Michael D.",
    title = "{Polarized image of equatorial emission in the Kerr geometry}",
    eprint = "2105.09440",
    archivePrefix = "arXiv",
    primaryClass = "gr-qc",
    doi = "10.1103/PhysRevD.104.044060",
    journal = {\prd},
    volume = "104",
    number = "4",
    pages = "044060",
    year = "2021"
}

@article{Wielgus:2022heh,
    author = "Wielgus, Maciek and Moscibrodzka, Monika and Vos, Jesse and Gelles, Zachary and Marti-Vidal, Ivan and Farah, Joseph and Marchili, Nicola and Goddi, Ciriaco and Messias, Hugo",
    title = "{Orbital motion near Sagittarius A* - Constraints from polarimetric ALMA observations}",
    eprint = "2209.09926",
    archivePrefix = "arXiv",
    primaryClass = "astro-ph.HE",
    doi = "10.1051/0004-6361/202244493",
    journal = {\aap},
    volume = "665",
    pages = "L6",
    year = "2022"
}

@article{Cardenas-Avendano:2022csp,
    author = "C{\'a}rdenas-Avenda{\~n}o, Alejandro and Lupsasca, Alexandru and Zhu, Hengrui",
    title = "{Adaptive analytical ray tracing of black hole photon rings}",
    eprint = "2211.07469",
    archivePrefix = "arXiv",
    primaryClass = "gr-qc",
    doi = "10.1103/PhysRevD.107.043030",
    journal = {\prd},
    volume = "107",
    number = "4",
    pages = "043030",
    year = "2023"
}

@article{Ricarte:2025aix,
    author = "Ricarte, Angelo and Conroy, Nicholas S. and Wielgus, Maciek and Palumbo, Daniel and Emami, Razieh and Chan, Chi-kwan",
    title = "{Dynamical Inference from Polarized Light Curves of Sagittarius A*}",
    eprint = "2504.01114",
    archivePrefix = "arXiv",
    primaryClass = "astro-ph.HE",
    doi = "10.3847/1538-4357/add729",
    journal = {\apj},
    volume = "987",
    number = "2",
    pages = "152",
    month = "7",
    year = "2025"
}

@article{Walker:1970un,
    author = "Walker, M. and Penrose, R.",
    title = "{On quadratic first integrals of the geodesic equations for type [22] spacetimes}",
    doi = "10.1007/BF01649445",
    journal = {CMaPh},
    volume = "18",
    pages = "265--274",
    year = "1970"
}

@article{Levis:2023tpb,
    author = "Levis, Aviad and Chael, Andrew A. and Bouman, Katherine L. and Wielgus, Maciek and Srinivasan, Pratul P.",
    title = "{Orbital polarimetric tomography of a flare near the Sagittarius A$^{*}$ supermassive black hole}",
    eprint = "2310.07687",
    archivePrefix = "arXiv",
    primaryClass = "astro-ph.HE",
    doi = "10.1038/s41550-024-02238-3",
    journal = {NatAs},
    volume = "8",
    number = "6",
    pages = "765--773",
    year = "2024"
}

@misc{WakeHPC,
  doi       = {10.57682/G13Z-2362},
  url       = {https://hpc.wfu.edu},
  author    = {{Information Systems and Wake Forest University}},
  title     = {{WFU High Performance Computing Facility}},
  publisher = {Wake Forest University},
  year      = {2021}
}

@article{Yfantis:2024eab,
    author = "Yfantis, A. I. and Wielgus, M. and Mo{\'s}cibrodzka, M. A.",
    title = "{Hot spots around Sagittarius A* - Joint fits to astrometry and polarimetry}",
    eprint = "2408.07120",
    archivePrefix = "arXiv",
    primaryClass = "astro-ph.HE",
    doi = "10.1051/0004-6361/202451884",
    journal = {\aap},
    volume = "691",
    pages = "A327",
    year = "2024"
}

@article{Yfantis:2023wsp,
    author = "Yfantis, A. I. and Mo{\'s}cibrodzka, M. A. and Wielgus, M. and Vos, J. T. and Jimenez-Rosales, A.",
    title = "{Fitting the light curves of Sagittarius A* with a hot-spot model - Bayesian modeling of QU loops in the millimeter band}",
    eprint = "2310.07762",
    archivePrefix = "arXiv",
    primaryClass = "astro-ph.HE",
    doi = "10.1051/0004-6361/202348230",
    journal = {\aap},
    volume = "685",
    pages = "A142",
    year = "2024"
}

@article{Vincent:2023sbw,
    author = "Vincent, F. H. and Wielgus, M. and Aimar, N. and Paumard, T. and Perrin, G.",
    title = "{Polarized signatures of orbiting hot spots: Special relativity impact and probe of spacetime curvature}",
    eprint = "2309.10053",
    archivePrefix = "arXiv",
    primaryClass = "astro-ph.HE",
    doi = "10.1051/0004-6361/202348016",
    journal = {\aap},
    volume = "684",
    pages = "A194",
    year = "2024"
}

@article{Emami:2022ydq,
    author = "Emami, Razieh and others",
    title = "{Tracing Hot Spot Motion in Sagittarius A* Using the Next-Generation Event Horizon Telescope (ngEHT)}",
    eprint = "2211.06773",
    archivePrefix = "arXiv",
    primaryClass = "astro-ph.GA",
    doi = "10.3390/galaxies11010023",
    journal = {Galax},
    volume = "11",
    number = "1",
    pages = "23",
    year = "2023"
}

@article{Vos:2022yij,
    author = "Vos, Jesse and Moscibrodzka, Monika and Wielgus, Maciek",
    title = "{Polarimetric signatures of hot spots in black hole accretion flows}",
    eprint = "2209.09931",
    archivePrefix = "arXiv",
    primaryClass = "astro-ph.HE",
    doi = "10.1051/0004-6361/202244840",
    journal = {\aap},
    volume = "668",
    pages = "A185",
    year = "2022"
}

@article{Ball:2020jup,
    author = {Ball, David and {\"O}zel, Feryal and Christian, Pierre and Chan, Chi-Kwan and Psaltis, Dimitrios},
    title = "{A Plasmoid model for the Sgr A* Flares Observed With Gravity and CHANDRA}",
    eprint = "2005.14251",
    archivePrefix = "arXiv",
    primaryClass = "astro-ph.HE",
    doi = "10.3847/1538-4357/abf8ae",
    journal = {\apj},
    volume = "917",
    number = "1",
    pages = "8",
    year = "2021"
}

@book{Rybicki:2004hfl,
    author = "Rybicki, George B. and Lightman, Alan P.",
    title = "{Radiative Processes in Astrophysics}",
    doi = "10.1002/9783527618170",
    isbn = "978-0-471-82759-7, 978-3-527-61817-0",
    publisher = "Wiley-VCH",
    year = "2004"
}

@article{EventHorizonTelescope:2021btj,
    author = "Narayan, Ramesh and Palumbo, Daniel C. M. and Johnson, Michael D. and others",
    collaboration = "Event Horizon Telescope",
    title = "{The Polarized Image of a Synchrotron-emitting Ring of Gas Orbiting a Black Hole}",
    eprint = "2105.01804",
    archivePrefix = "arXiv",
    primaryClass = "astro-ph.HE",
    reportNumber = "FERMILAB-PUB-21-848-PPD",
    doi = "10.3847/1538-4357/abf117",
    journal = {\apj},
    volume = "912",
    number = "1",
    pages = "35",
    year = "2021"
}

@article{GRAVITY:2020lpa,
    author = {Baub{\"o}ck, M. and others},
    collaboration = "GRAVITY",
    title = "{Modeling the orbital motion of Sgr A*{\textquoteright}s near-infrared flares}",
    eprint = "2002.08374",
    archivePrefix = "arXiv",
    primaryClass = "astro-ph.HE",
    doi = "10.1051/0004-6361/201937233",
    journal = {\aap},
    volume = "635",
    pages = "A143",
    year = "2020"
}

@article{GRAVITY:2023avo,
    author = "Abuter, R. and others",
    collaboration = "GRAVITY",
    title = "{Polarimetry and astrometry of NIR flares as event horizon scale, dynamical probes for the mass of Sgr A*}",
    eprint = "2307.11821",
    archivePrefix = "arXiv",
    primaryClass = "astro-ph.GA",
    doi = "10.1051/0004-6361/202347416",
    journal = {\aap},
    volume = "677",
    pages = "L10",
    year = "2023"
}

@article{EventHorizonTelescope:2022apq,
    author = "Akiyama, Kazunori and others",
    collaboration = "Event Horizon Telescope",
    title = "{First Sagittarius A* Event Horizon Telescope Results. II. EHT and Multiwavelength Observations, Data Processing, and Calibration}",
    eprint = "2311.08679",
    archivePrefix = "arXiv",
    primaryClass = "astro-ph.HE",
    reportNumber = "FERMILAB-PUB-22-418-PPD",
    doi = "10.3847/2041-8213/ac6675",
    journal = {\apjl},
    volume = "930",
    number = "2",
    pages = "L13",
    year = "2022"
}

@article{Haggard:2019mro,
    author = "Haggard, Daryl and others",
    title = "{Chandra Spectral and Timing Analysis of Sgr A*'s Brightest X-ray Flares}",
    eprint = "1908.01781",
    archivePrefix = "arXiv",
    primaryClass = "astro-ph.HE",
    doi = "10.3847/1538-4357/ab4a7f",
    journal = {\apj},
    volume = "886",
    number = "2",
    pages = "96",
    year = "2019"
}

@article{Najafi-Ziyazi:2023oil,
    author = "Najafi-Ziyazi, Mahdi and Davelaar, Jordy and Mizuno, Yosuke and Porth, Oliver",
    title = "{Flares in the Galactic centre {\textendash} II. Polarization signatures of flares at mm-wavelengths}",
    eprint = "2308.16740",
    archivePrefix = "arXiv",
    primaryClass = "astro-ph.HE",
    doi = "10.1093/mnras/stae1343",
    journal = {\mnras},
    volume = "531",
    number = "4",
    pages = "3961--3972",
    year = "2024"
}

@article{Porth:2020txf,
    author = "Porth, O. and Mizuno, Y. and Younsi, Z. and Fromm, C. M.",
    title = "{Flares in the Galactic Centre {\textendash} I. Orbiting flux tubes in magnetically arrested black hole accretion discs}",
    eprint = "2006.03658",
    archivePrefix = "arXiv",
    primaryClass = "astro-ph.HE",
    doi = "10.1093/mnras/stab163",
    journal = {\mnras},
    volume = "502",
    number = "2",
    pages = "2023--2032",
    year = "2021"
}

@article{EventHorizonTelescope:2022ago,
    author = "Wielgus, Maciek and others",
    title = "{Millimeter Light Curves of Sagittarius A* Observed during the 2017 Event Horizon Telescope Campaign}",
    eprint = "2207.06829",
    archivePrefix = "arXiv",
    primaryClass = "astro-ph.HE",
    doi = "10.3847/2041-8213/ac6428",
    journal = {\apjl},
    volume = "930",
    number = "2",
    pages = "L19",
    year = "2022"
}

@article{Ripperda:2021zpn,
    author = "Ripperda, Bart and Liska, Matthew and Chatterjee, Koushik and Musoke, Gibwa and Philippov, Alexander A. and Markoff, Sera B. and Tchekhovskoy, Alexander and Younsi, Ziri",
    title = "{Black Hole Flares: Ejection of Accreted Magnetic Flux through 3D Plasmoid-mediated Reconnection}",
    eprint = "2109.15115",
    archivePrefix = "arXiv",
    primaryClass = "astro-ph.HE",
    doi = "10.3847/2041-8213/ac46a1",
    journal = {\apjl},
    volume = "924",
    number = "2",
    pages = "L32",
    year = "2022"
}

@article{Lupsasca:2024xhq,
    author = "Lupsasca, Alexandru and C{\'a}rdenas-Avenda{\~n}o, Alejandro and Palumbo, Daniel C. M. and Johnson, Michael D. and Gralla, Samuel E. and Marrone, Daniel P. and Galison, Peter and Tiede, Paul and Keeble, Lennox",
    title = "{The Black Hole Explorer: photon ring science, detection, and shape measurement}",
    eprint = "2406.09498",
    archivePrefix = "arXiv",
    primaryClass = "gr-qc",
    doi = "10.1117/12.3019437",
    journal = {\procspie},
    volume = "13092",
    pages = "130926Q",
    year = "2024"
}

@article{Pu:2016qak,
    author = "Pu, Hung-Yi and Akiyama, Kazunori and Asada, Keiichi",
    title = "{The Effects of Accretion Flow Dynamics on the Black Hole Shadow of Sagittarius A$^{*}$}",
    eprint = "1608.03035",
    archivePrefix = "arXiv",
    primaryClass = "astro-ph.HE",
    doi = "10.3847/0004-637X/831/1/4",
    journal = {\apj},
    volume = "831",
    number = "1",
    pages = "4",
    year = "2016"
}

@article{Broderick:2005jj,
    author = "Broderick, Avery E. and Loeb, Abraham",
    title = "{Imaging optically-thin hot spots near the black hole horizon of sgr a* at radio and near-infrared wavelengths}",
    eprint = "astro-ph/0509237",
    archivePrefix = "arXiv",
    doi = "10.1111/j.1365-2966.2006.10152.x",
    journal = {\mnras},
    volume = "367",
    pages = "905--916",
    year = "2006"
}

@article{Broderick:2005my,
    author = "Broderick, Avery E. and Loeb, Abraham",
    title = "{Imaging bright spots in the accretion flow near the black hole horizon of Sgr A*}",
    eprint = "astro-ph/0506433",
    archivePrefix = "arXiv",
    doi = "10.1111/j.1365-2966.2005.09458.x",
    journal = {\mnras},
    volume = "363",
    pages = "353--362",
    year = "2005"
}

@article{Mossoux:2020ddc,
    author = "Mossoux, E. and Finociety, B. and Beckers, J. -M. and Vincent, F. H.",
    title = "{Continuation of the X-ray monitoring of Sgr A*: the increase in bright flaring rate confirmed}",
    eprint = "2003.06191",
    archivePrefix = "arXiv",
    primaryClass = "astro-ph.HE",
    doi = "10.1051/0004-6361/201937136",
    journal = {\aap},
    volume = "636",
    pages = "A25",
    year = "2020"
}

@article{Zamaninasab:2009df,
    author = "Zamaninasab, M. and others",
    title = "{Near infrared flares of Sagittarius A*: Importance of near infrared polarimetry}",
    eprint = "0911.4659",
    archivePrefix = "arXiv",
    primaryClass = "astro-ph.GA",
    doi = "10.1051/0004-6361/200912473",
    journal = {\aap},
    volume = "510",
    pages = "A3",
    year = "2010"
}

@ARTICLE{1992A&A...257..531K,
       author = {{Karas}, V. and {Bao}, G.},
        title = "{On the light curve of an orbiting spot.}",
      journal = {\aap},
         year = 1992,
        month = apr,
       volume = {257},
        pages = {531-533},
       adsurl = {https://ui.adsabs.harvard.edu/abs/1992A&A...257..531K}
}

@article{Wielgus:2023akf,
  author = {Wielgus, Maciek and Issaoun, Sara and Marti-Vidal, Ivan and Emami, Razieh and Moscibrodzka, Monika and Brinkerink, Christiaan D. and Goddi, Ciriaco and Fomalont, Ed},
  title = {{The internal Faraday screen of Sagittarius A*}},
  journal = {\aap},
  volume = {682},
  pages = {A97},
  year = {2024},
  eprint = {2308.11712},
  archivePrefix = {arXiv},
  primaryClass = {astro-ph.HE},
  doi = {10.1051/0004-6361/202347772}
}

@article{Tlemissov:2026kuu,
       author = {{Tlemissov}, Abylaikhan and {Tursunov}, Arman and {Wielgus}, Maciek},
        title = "{Polarized emission of orbiting hot spots near Sagittarius A*: Effects of electromagnetic interaction}",
      journal = {\aap},
         year = 2026,
       volume = {712},
          eid = {A104},
        pages = {A104},
          doi = {10.1051/0004-6361/202557429},
archivePrefix = {arXiv},
       eprint = {2606.18107},
 primaryClass = {astro-ph.HE},
       adsurl = {https://ui.adsabs.harvard.edu/abs/2026A&A...712A.104T}
}

@article{Ruales:2026xjb,
  title = {Polarization signatures of inspiraling hotspots around Kerr black holes},
  author = {Ruales, Pablo and Gates, Delilah E. A. and C{\'a}rdenas-Avenda{\~n}o, Alejandro},
  journal = {\prd},
  volume = {113},
  number = {10},
  pages = {103030},
  year = {2026},
  month = {May},
  eprint = {2602.09102},
  archivePrefix = {arXiv},
  doi = {10.1103/65fp-f288}
}

@article{2026A&A...706A.323P,
    author = "Pe{\~n}a R., Pablo A. and Jenkins, James S.",
    title = "{Closing the evidence gap: reddemcee, a fast adaptive parallel tempering sampler}",
    eprint = "2509.24870",
    archivePrefix = "arXiv",
    primaryClass = "astro-ph.IM",
    doi = "10.1051/0004-6361/202556609",
    journal = {\aap},
    volume = "706",
    pages = "A323",
    year = "2026"
}

@article{Rojas-Paternina:2026nsq,
    author = "Rojas-Paternina, Daniel and C{\'a}rdenas-Avenda{\~n}o, Alejandro",
    title = "{Light propagation prescriptions for black hole movies}",
    eprint = "2605.12659",
    archivePrefix = "arXiv",
    primaryClass = "astro-ph.HE",
    doi = "10.1103/nsg2-hy9p",
    journal = {\prd},
    volume = "114",
    number = "2",
    pages = "023031",
    year = "2026"
}

@article{Baganoff:2001kw,
    author = "Baganoff, F. K. and others",
    title = "{Rapid X-ray flaring from the direction of the supermassive black hole at the galactic centre}",
    eprint = "astro-ph/0109367",
    archivePrefix = "arXiv",
    primaryClass = "astro-ph",
    doi = "10.1038/35092510",
    journal = {\nat},
    volume = "413",
    pages = "45--48",
    year = "2001"
}

@article{Genzel:2003as,
    author = "Genzel, R. and Schoedel, R. and Ott, T. and Eckart, A. and Alexander, T. and Lacombe, F. and Rouan, D. and Aschenbach, B.",
    title = "{Near-infrared flares from accreting gas around the supermassive black hole at the Galactic Centre}",
    eprint = "astro-ph/0310821",
    archivePrefix = "arXiv",
    primaryClass = "astro-ph",
    doi = "10.1038/nature02065",
    journal = {\nat},
    volume = "425",
    pages = "934--937",
    year = "2003"
}

@article{Abuter:2018uum,
    author = {Abuter, R. and others},
    collaboration = "GRAVITY",
    title = "{Detection of orbital motions near the last stable circular orbit of the massive black hole SgrA*}",
    eprint = "1810.12641",
    archivePrefix = "arXiv",
    primaryClass = "astro-ph.GA",
    doi = "10.1051/0004-6361/201834294",
    journal = {\aap},
    volume = "618",
    pages = "L10",
    year = "2018"
}

@article{Johnson:2024ttr,
    author = "Johnson, Michael D. and others",
    title = "{The Black Hole Explorer: Motivation and Vision}",
    eprint = "2406.12917",
    archivePrefix = "arXiv",
    primaryClass = "astro-ph.IM",
    doi = "10.1117/12.3019835",
    journal = {\procspie},
    volume = "13092",
    pages = "130922D",
    year = "2024"
}

@article{Narayan:1994xi,
    author = "Narayan, Ramesh and Yi, In-su",
    title = "{Advection dominated accretion: A Selfsimilar solution}",
    eprint = "astro-ph/9403052",
    archivePrefix = "arXiv",
    reportNumber = "CFA-3809",
    doi = "10.1086/187381",
    journal = {\apjl},
    volume = "428",
    pages = "L13",
    year = "1994"
}

@article{GRAVITY:2021xju,
    author = "Abuter, R. and others",
    collaboration = "GRAVITY",
    title = "{Mass distribution in the Galactic Center based on interferometric astrometry of multiple stellar orbits}",
    eprint = "2112.07478",
    archivePrefix = "arXiv",
    primaryClass = "astro-ph.GA",
    doi = "10.1051/0004-6361/202142465",
    journal = {\aap},
    volume = "657",
    pages = "L12",
    year = "2022"
}

@article{Johnson:2023ynn,
    author = "Johnson, Michael D. and others",
    title = "{Key Science Goals for the Next-Generation Event Horizon Telescope}",
    eprint = "2304.11188",
    archivePrefix = "arXiv",
    primaryClass = "astro-ph.HE",
    doi = "10.3390/galaxies11030061",
    journal = {Galax},
    volume = "11",
    number = "3",
    pages = "61",
    year = "2023"
}

@article{Cardenas-Avendano:2026bow,
    author = "C{\'a}rdenas-Avenda{\~n}o, Alejandro and Rubiera-Garcia, Diego and Vincent, Frederic H.",
    title = "{A Four-Dimensional Gaussian Random Field Generator for Modeling Spatiotemporal Variability in Astrophysical Sources}",
    journal = {arXiv e-prints},
    pages = {arXiv:2607.16576},
    doi = {10.48550/arXiv.2607.16576},
    eprint = "2607.16576",
    archivePrefix = "arXiv",
    primaryClass = "astro-ph.HE",
    month = "7",
    year = "2026"
}

@ARTICLE{2023ApJ...950...38E,
       author = {{Emami}, Razieh and {Ricarte}, Angelo and {Wong}, George N. and {Palumbo}, Daniel and {Chang}, Dominic and {Doeleman}, Sheperd S. and {Broderick}, Avery E. and {Narayan}, Ramesh and {Wielgus}, Maciek and {Blackburn}, Lindy and {Prather}, Ben S. and {Chael}, Andrew A. and {Anantua}, Richard and {Chatterjee}, Koushik and {Marti-Vidal}, Ivan and {G{\'o}mez}, Jose L. and {Akiyama}, Kazunori and {Liska}, Matthew and {Hernquist}, Lars and {Tremblay}, Grant and {Vogelsberger}, Mark and {Alcock}, Charles and {Smith}, Randall and {Steiner}, James and {Tiede}, Paul and {Roelofs}, Freek},
        title = "{Unraveling Twisty Linear Polarization Morphologies in Black Hole Images}",
      journal = {\apj},
         year = 2023,
        month = jun,
       volume = {950},
       number = {1},
          eid = {38},
        pages = {38},
          doi = {10.3847/1538-4357/acc8cd},
archivePrefix = {arXiv},
       eprint = {2210.01218},
 primaryClass = {astro-ph.GA},
       adsurl = {https://ui.adsabs.harvard.edu/abs/2023ApJ...950...38E}
}

@inproceedings{Wielgus:2026qft,
    author = "Wielgus, Maciek and Yfantis, Aristomenis",
    title = "{Dissecting the variability of Sagittarius A* in the orbiting hotspot framework}",
    booktitle = "{IAU Symposium 405: Traversing the Galactic Center in Space and Time}",
    eprint = "2609.28844",
    archivePrefix = "arXiv",
    primaryClass = "astro-ph.HE",
    month = "9",
    year = "2026"
}
\bibliographystyle{aasjournalv7}

\end{document}